# Symmetry recovery of exceptional points and their dynamical encircling in a two-state system

Xu-Lin Zhang,[1,2] Shubo Wang,[1] Wen-Jie Chen,[1] Bo Hou,[1,3] and C. T. Chan[1,*]

*[1]Department of Physics and Institute for Advanced Study, The Hong Kong University of Science and Technology, Clear Water Bay, Hong Kong, China*
*[2]State Key Laboratory on Integrated Optoelectronics, College of Electronic Science and Engineering, Jilin University, Changchun, China*
*[3]College of Physics, Optoelectronics and Energy & Collaborative Innovation Center of Suzhou Nano Science and Technology, Soochow University, Suzhou, China*

[*]Corresponding author: phchan@ust.hk

Exceptional points are degeneracies in non-Hermitian systems. A two-state system with parity-time (*PT*) symmetry usually has only one exceptional point beyond which the eigenmodes are *PT*-symmetry broken. The so-called symmetry recovery, i.e., eigenmodes become *PT*-symmetric again, typically occurs in multi-state systems. Here we show that a two-state ferromagnetic waveguide system can have an exceptional point and a subsequent symmetry recovery due to the presence of accidental degeneracy points when the system is lossless. By introducing a parameter space where both exceptional points reside, we designed a system in which the trajectory in the parameter space can be controlled in situ using an adiabatically tunable external field, allowing us to explore the topological and chiral character of the system by encircling zero, one or two exceptional points. We performed microwave experiments to demonstrate the presence of the exceptional point, symmetry recovery, and the effects arising from their dynamical encircling.

Non-Hermitian systems can possess real eigenvalues if their complex potential has parity-time (*PT*) symmetry[1]. The system undergoes a transition at an "exceptional point" (EP) when the degree of non-Hermiticity is increased[2-4], after which the real parts of the eigenvalues coalesce while the imaginary parts bifurcate. In optics, *PT* symmetry can be realized if the system possesses the symmetry $\varepsilon(x) = \varepsilon^*(-x)$ and $\mu(x) = \mu^*(-x)$[5]. Although it is difficult to realize optical gain experimentally, EPs can also be found in passive non-Hermitian systems with asymmetric loss, which can be viewed as *PT*-symmetric systems with a background of uniform loss[6]. The unique physics of EPs has given rise to many interesting phenomena and applications such as unusual beam dynamics[7,8], lasing effects[9-12], unidirectional transmissions[13,14], asymmetric mode switching[15], and others[16-23]. EPs are typically manipulated by tuning gain and loss[5,6]. When gain and loss are increased relative to coupling strength, there is one EP for a two-state linear system, and the *PT*-symmetric state cannot be recovered once the system enters the symmetry-broken phase. However, the so-called symmetry recovery (i.e., eigenmodes re-enter the *PT*-symmetric phase) can occur if the system has more than two states interacting[21,24,25] or if the system is nonlinear[26,27]. The topological structure of EPs has also attracted immense attention[28,29] since intriguing behavior such as a state flip could occur if an EP is encircled adiabatically. These effects were reported only recently in microwave[15] and optomechanical[30] experiments via a full dynamical encircling of the EP, showing that the output state is determined solely by the sense of rotation in the parameter space, regardless of the input state. A full analytical model[31] was established to explain this chiral nature of the dynamics, which can also be related to the Stokes phenomenon of asymptotics[32,33] and stability aspects[34]. Although a full dynamical encircling of one EP was realized experimentally, the encircling loop could not be changed once the sample was fabricated[15]. It would be even more interesting if we could dynamically tune the encircling loop and encircle more than one EP as it would reveal more insights into the topological structure of the EPs and enrich their applications.

In this work, we report the first experimental observation of an EP and a subsequent symmetry recovery in a two-state system as well as a dynamical encircling of more than one EP. We first investigate a *PT*-symmetric coupled ferromagnetic waveguide system in the presence of a bias magnetic field. The system possesses multiple degeneracy points (DPs), which are diabolical points due to the absence of mode couplings at specific field strengths. Each DP becomes a pair of EPs of opposite chirality when gain and loss are introduced. As a result, the system carries multiple EPs and exhibits symmetry recovery behaviors in a dynamical process when the external field is tuned monotonously. Using a pair of passive waveguides, we experimentally demonstrate the presence of a pair of EPs spawn from a single DP at microwave frequencies by measuring the transmission spectra and field distributions. We further design a coupled system in which the bias field and the width of one of the waveguides vary along the waveguiding direction. A dynamical encircling loop can be created in the parameter space where the two EPs (i.e., one EP and a subsequent symmetry recovery) reside. Remarkably, the trajectory in the parameter space can be controlled in situ using a tunable external field, and as such we can choose to encircle zero, one or even two EPs to explore the parameter space topology without changing or moving the sample. Experimental measurements on the mode symmetry demonstrate the effects arising from the dynamical encircling of the EPs.

## Results

**Exceptional point and symmetry recovery in a two-state system.** We start by showing the physical mechanism underlying the presence of multiple EPs for a two-state system. Figure 1a illustrates the $PT$-symmetric coupled ferromagnetic waveguide system with balanced gain and loss [$\varepsilon_1 = 12.3\text{-}i\gamma$, $\varepsilon_2 = 12.3+i\gamma$]. Static magnetic fields are applied to the two waveguides along opposite directions, inducing a diagonal term $\mu_b$ and an off-diagonal term $\pm i\chi$ of the permeability tensors[35]. The background material is assumed to have $\varepsilon_3 = 12$ and $\mu_3 = 1$. For simplicity's sake, we first set $\mu_b = 1$, which does not affect the underlying physics. We calculate the dispersions of the weakly guided modes in this paraxial waveguide system without gain or loss ($\gamma = 0$) using COMSOL[36]. Figure 1b shows the effective mode index, defined as $n_{eff} = \beta_z/k_0$, with $\beta_z$ and $k_0$ denoting the mode propagation constant and vacuum wave number respectively, for the fourth pair of symmetric and anti-symmetric modes as a function of $\chi$ (solid lines). Here, $\chi$ depends on the external magnetic field, as will be stipulated later. For comparison, we show the dispersion of the fourth mode in a single waveguide (dashed line), which crosses the coupled symmetric/anti-symmetric modes at two DPs. The mode degeneracy phenomenon has been reported in ferromagnetic waveguides in the presence of an external magnetic field due to the absence of mode couplings[37]. To substantiate this point, we define a mode coupling coefficient $\eta = \iint \left( \mathbf{E}_1 \mathbf{D}_2^* + \mathbf{H}_1 \mathbf{B}_2^* \right) d\sigma$ to represent the coupling strength between the two waveguides, where the subscript $j$ corresponds to the uncoupled eigenfield in the system with only waveguide $j$. Figure 1c shows the coupling coefficient between the uncoupled fourth set of modes. We find two regions with vanishing coupling strengths that match well with the two DPs in Fig. 1b. The absence of mode couplings in this system can be attributed to the mode symmetry transition from linear polarization to elliptical polarization induced by the transverse bias magnetic fields (see Supplementary Figs 1-3 and Supplementary Note 1 for a detailed discussion on the origin of the DPs as well as the behaviors of modes of other orders).

The coupled waveguide system without gain or loss can be described by a Hamiltonian of the form $H = \begin{bmatrix} \beta_0 & \kappa \\ \kappa^* & \beta_0 \end{bmatrix}$, where $\beta_0$ is the propagation constant of an uncoupled mode in the single waveguide and $\kappa$ denotes the couplings between the two uncoupled modes. We use the numerical results in Fig. 1b to fit $\kappa$ as a function of $\chi$. Figure 2a plots the fitted curve (solid line), showing that $\kappa$ displays a similar dependence on $\chi$ as the coupling coefficient $\eta$ in Fig. 1c. We then introduce gain and loss into the system. We show three cases of gain and loss ($\gamma = 3\times10^{-4}$, $4.9\times10^{-4}$, $2\times10^{-3}$) in Fig. 2b-d respectively, where the effective mode index becomes complex. A *PT*-broken phase emerges at an EP when the gain or loss parameter exceeds the coupling. As the coupling is vanishing at the DPs, we expect that even a very small $\gamma$ can give rise to EPs. This is indeed the case, as shown in Fig. 2b, where a small $\gamma = 3\times10^{-4}$ is sufficient to turn each DP into a pair of EPs. For ordinary two-level systems, there is only one EP and the *PT*-symmetry of eigenmodes remains broken beyond that EP. However, as $\chi$ increases, our system first enters a *PT*-broken phase but re-enters a *PT*-symmetric phase before becoming *PT*-broken again. *PT*-symmetry recovery has been reported in multi-state systems[21,24,25], but it occurs here in a two-state system due to the non-monotonic couplings. When the gain or loss reaches a particular threshold ($\gamma = 4.9\times10^{-4}$), the right-hand EP of the left bubble and the left-hand EP of the right bubble merge to form a DP, as shown in Fig. 2c. The real eigenvalues cross linearly for ordinary diabolical points, but in the case of this DP it is the imaginary parts that cross linearly. As $\gamma$ increases further, the two *PT*-broken phase regions merge as shown in Fig. 2d. The corresponding chirality of the EPs is shown in Fig. 2b-d. To better understand the variation in the eigenmode behaviors, we calculate the magnitude of the imaginary part of the propagation constant (defined as $\beta_r \pm i\beta_i$) in a single waveguide for the above three cases of gain and loss and plot it in Fig. 2a (see dashed lines). The regimes corresponding to $|\kappa| < \beta_i$ coincide with the numerically obtained *PT*-broken phase regimes in Fig. 2b-d.

In fact, a system with asymmetric losses (rather than exact *PT* symmetry) can exhibit the aforementioned effects. We design a non-Hermitian passive system consisting of a pair of yttrium iron garnet (YIG) dielectric waveguides working at microwave frequencies. The background material is assumed to be air. Microwave absorbers are attached to the side of YIG waveguide 2 (red region in Fig. 3a) to introduce asymmetric losses into the system, given that the intrinsic dielectric loss of YIG is negligible. We apply a bias magnetic field along the negative $x$-axis. The eigenfield distributions move towards the $+y$ interface due to the field displacement effect when the transverse bias magnetic field is perpendicular to the RF magnetic field[35]. The EP and symmetry recovery can also appear in this configuration due to the field-induced variation in mode couplings. We first theoretically analyze the eigenmodes supported in this system. The permeability tensor of YIG is modeled with $\mu_b = 1 + \omega_m \omega_0 / \left(\omega_0^2 - \omega^2\right)$ and $\chi = \omega_m \omega / \left(\omega_0^2 - \omega^2\right)$, where $\omega_0 = \mu_0 \gamma_R H_0$ is determined by the gyromagnetic ratio $\gamma_R$ and bias magnetic field $H_0$, and $\omega_m = \mu_0 \gamma_R M$ is determined by the magnetization $M$, which is measured experimentally (Supplementary Fig. 4). Here, permeability losses can be ignored because the system operates far away from the gyromagnetic resonance. Figure 3b shows the calculated effective mode index of the coupled YIG waveguides at 9.5 GHz as a function of the bias magnetic field. In the simulation, the permittivity of YIG and that of the microwave absorber are chosen as ~15.2[38] and ~4+15$i$ (Supplementary Fig. 4) respectively. We find a region (0.02 T~0.16 T) where the real parts of the two modes almost coalesce while the corresponding imaginary parts first repel and then attract each other again. This is a typical feature of an EP associated with symmetry recovery, although here the real/imaginary parts are separated in the broken/symmetric phase region due to the mode detuning induced by the broken symmetry of the two waveguides (i.e., only one waveguide has a microwave absorber attached). In other words, the system is very close to the ‘exact’ EP and symmetry recovery point (Supplementary Fig. 5). The inset of Fig. 3b shows the emergence of a DP when the microwave absorber is removed. This DP spawns a pair of EPs of opposite chirality when asymmetric losses are introduced and symmetry recovery naturally arises. The power flow distributions of the eigenmodes are also shown in the inset of Fig. 3b, where the field displacement effect can be found. The symmetry of the eigenmode (*PT*-symmetric or *PT*-broken) is also evident.

We then perform microwave experiments to implement the theoretical design. A photograph of the experimental setup is shown in Fig. 4a, where a pair of YIG waveguides is placed inside a vibrating sample magnetometer (VSM), which can provide a quasi-uniform bias magnetic field along the negative *x*-axis in the area of interest. The transmission spectra of the system are measured to identify the EP as well as the symmetry recovery behavior, since we expect the transmission to be enhanced in the broken phase region provided that both eigenmodes are excited efficiently[6]. The measured transmission spectra for different magnetic field strengths and frequencies are plotted in Fig. 4b. We observe that for each frequency the transmission is dramatically enhanced under certain bias fields. The enhanced transmission regime corresponds to the symmetry-broken phase, in which the wave can travel a longer distance in the waveguide without the absorber. For example, we observe enhanced transmission at 9.5 GHz mainly between 0.02 T and 0.16 T, highly consistent with the broken phase region predicted in Fig. 3b. This enhanced region shifts to higher external fields at lower frequencies, due to the shift in the DPs in the lossless system. The black dashed line in Fig. 4b marks the calculated DPs and lies almost at the center of the experimental broken phase region, indicating the EPs are indeed spawned from the DPs.

To investigate the underlying physics, we measure the electric field intensity on the surface of each waveguide along the waveguiding direction (see *z*-axis in Fig. 4a). The experimental results at 9.5 GHz for different bias fields are shown in Fig. 4c,d, respectively, for waveguide 1 and waveguide 2. The transmission enhancement regime at 9.5 GHz estimated from Fig. 4b is marked by the two grey dashed lines, which partition the regime into two symmetric phases and one broken phase. In the symmetric phase regime, both the symmetric and anti-symmetric modes experience considerable losses, suggesting that the electric field intensity decays exponentially along the direction of propagation in both waveguide 1 (Fig. 4c) and waveguide 2 (Fig. 4d). In the broken phase region, however, the two eigenmodes propagate individually in the two waveguides (see the power flow patterns in the inset of Fig. 3b). As a result, the wave in waveguide 1 can travel a longer distance since the mode losses are small (Fig.

4c), whereas the wave in waveguide 2 experiences higher losses because of the attached absorber (Fig. 4d). The transmission of the whole system is then enhanced as a result of the longer transport distance in the lossless waveguide 1. The large contrast between the field distributions in the two waveguides convincingly demonstrates the presence of the symmetric phase and the broken phase. More results of the control experiments are given in Supplementary Figs 6-9 and Supplementary Note 2.

We perform numerical simulations to support the experimental results. To calculate the transmission spectra, we place a line current source (white lines in Fig. 3d) near the edge of the lossy waveguide to mimic the antenna in experiments. The power flow $S_z$ at the output surface is then integrated to provide the transmission. Figure 3c shows the calculated transmission spectrum of a 200 mm system at 9.5 GHz, which can reproduce the salient features observed in experiments. The transmission exhibits an oscillating behavior in both experiments and simulations due to the Fabry-Perot resonance associated with the change in the effective mode index as the bias magnetic field varies. Increasing the waveguide length further to 400 mm in the simulation can double the resonant peak number (see Fig. 3c), confirming the presence of the Fabry-Perot effect. We show in Fig. 3d the calculated power flow distributions for four cases marked (i)-(iv) in Fig. 3c to illustrate the mode behaviors in the symmetric phase regime ((i) and (iv)) and the broken phase regime ((ii) and (iii)). The simulated field distributions coincide with the experimental results in Fig. 4c,d. We also note from Fig. 3c that beyond the broken phase region, the transmission again increases when the field exceeds 0.2 T (also see Fig. 4b for experimental results). This is due to the overall decrease in mode losses (see the imaginary part in Fig. 3b) as field strength increases because a stronger external field can displace the eigenfield from the absorber (see the field patterns in Fig. 3b). A further increase in the external field drives the two eigenmodes close to the cut-off (see the real part in Fig. 3b; also refer to the white dashed line in Fig. 4b) as $\mu_b$ approaches zero[35]. As a result, the transmission shows a sudden drop. All these numerical results are consistent with the experimental results, demonstrating the presence of the EP as well as the symmetry recovery.

**A dynamical encircling of the exceptional point and the symmetry recovery point.** The bias magnetic field in this system can be tuned adiabatically and continuously, making our system a good platform on which to study the topological properties associated with the EPs[15,30,31,34,39]. Forming an encircling loop in a parameter space requires two system parameters that can be changed continuously in space or time. In addition to the bias field, we choose the width of YIG waveguide 2 as the second parameter which is determined by a scale factor $\alpha$ (see Fig. 5b for definition). We calculate the effective mode index for the system at 9 GHz as a function of both the scale factor of waveguide 2 and the bias field (see the caption of Fig. 5b for detailed system parameters). The obtained Riemann surface for the lossless system is shown in Fig. 5e where a DP emerges. When asymmetric loss is introduced, the DP spawns a pair of EPs, exhibiting a self-intersecting Riemann surface as shown in Fig. 5f (real part) and Fig. 5g (imaginary part). The white dashed line in Fig. 5f marks the 'exact' broken phase region where the real parts of the two eigenvalues coalesce (also refer to the side view). The two end points of this broken phase line are EPs, beyond which the Riemann surface splits. The topological structure of our system is more complex than previous ones[15,30,31,34,39] due to the presence of two EPs, allowing us to gain more physical insights into the topological properties associated with EPs.

Following the idea that a dynamical encircling of one EP can be realized in a modulated waveguide system[15,39], we design a coupled system in Fig. 5a with the bias field and the scale factor $\alpha$ varying continuously along the waveguiding direction (i.e., $z$-axis). Given the diameter of the VSM magnet (~200 mm), the length of the coupled system is set to 400 mm so that the external magnetic field almost vanishes at the two terminals of the system. The experimentally measured bias field distribution along the $z$-axis is plotted with circles in Fig. 5c where $B_m$ denotes the maximum field at the center of the waveguides. The field distribution is closely fitted using a sinusoidal function (solid line in Fig. 5c) for further numerical simulations. Note that in the aforementioned experimental demonstration of the EP and symmetry recovery (Fig. 4), the bias field distribution is also non-uniform. But we show in Supplementary Fig. 10 that the non-uniformity would not affect the underlying physics of that system which is only 200 mm in length. Besides the bias field, the scale factor $\alpha$ is designed to vary along the $z$-axis with a minimum of 0.875 at $z$ = 100 mm and a maximum of 1.125 at $z$ = 300 mm (Fig. 5d, also see Fig. 5a,b for the schematic diagram). The parameter space is then defined in Fig. 6a, where the positions of the two EPs are also marked. Note that the scattering through our designed system is analogous to a loop in the defined parameter space, with the starting/end point at $B = 0$ and $\alpha = 1$. Injections from the left ($z = 0$) and right side ($z = 400$ mm) of the waveguide system (see the schematic diagram in Fig. 5a) correspond to counter-clockwise and clockwise loops, respectively. Some examples are given in Fig. 6a, where the green, black, and yellow loops are generated by $B_m = 0$ T, 0.08 T, and 0.17 T, corresponding to a dynamical encircling of zero, one, and two EPs, respectively. The encircling loop in our system can be tuned in situ continuously since the loop size is determined by an adiabatically tunable parameter, $B_m$. This was not possible in previous experimental work[15], where the encircling loop was fixed once the samples were fabricated.

We perform numerical simulations to demonstrate the effects arising from the dynamical encircling of EPs. Four cases are considered. Cases I and II correspond to counter-clockwise loops whereas cases III and IV clockwise loops. The injection is a symmetric mode for cases I and III and an anti-symmetric mode for cases II and IV. The modal transmission intensities $T_{nm}$ ($T_{nm}^{'}$), defined as the transmission from mode $m$ into mode $n$ in a counter-clockwise (clockwise) loop, where the subscript 1 denotes the symmetric mode and 2 the anti-symmetric mode, are applied to reveal the behavior of mode switching when the EPs are encircled.

Figure 6b plots the calculated transmission intensities of the proposed system at 9 GHz as a function of $B_m$ for the four cases. We find in case I that within a specific range of $B_m$ (i.e., 0.03 T~0.13 T), the anti-symmetric mode dominates the output, corresponding to a state flip. Outside of this range, however, the output is mainly a symmetric mode which is the same as the injection. The three loops defined in Fig. 6a can be used as examples to interpret these features. Their trajectories of the eigenstate evolution on the Riemann surface are also drawn in Fig. 5f,g. At $B_m = 0$, the loop does not encircle any EP so that the eigenstate returns to itself, which can also be verified by the $H_y$ field distribution in Fig. 6c. The situation is different when $B_m = 0.08$ T since the bigger loop now encircles one EP. The topological structure of the system then allows the two eigenmodes to flip after the EP is encircled (see the black line in Fig. 5f), as long as the encircling takes place on the Riemann sheet with lower losses[15,31,34,40] (see the black line in Fig. 5g). The state flip from symmetric input to anti-symmetric output can be understood intuitively through the dynamical evolution of the $H_y$ field shown in Fig. 6d. The unique topological structure of our system allows us to encircle two EPs if we further increase the bias field to, e.g., $B_m = 0.17$ T. However, the eigenstate returns to itself again because the chirality of one EP cancels out that of the other EP as they are spawn from the same DP (see the yellow lines in Fig. 5f,g for dynamical evolutions). The field distribution in Fig. 6e also shows that the state returns to itself after looping around both EPs. This process is nearly adiabatic in the presence of non-Hermiticity, since the field profile at any position follows closely that of the instantaneous eigenstate on the Riemann surface. Therefore, the variation in the output mode symmetry with increasing bias field indeed reflects a change in the number of EPs encircled in the parameter space.

We study case II from the same viewpoint. The transmission intensities in Fig. 6b indicate that the mode symmetry stays the same when zero or two EPs are encircled, as in case I. However, the eigenstate still returns to itself even if the loop encircles one EP, as demonstrated by the field evolution with $B_m = 0.08$ T in Fig. 6f. This behavior is due to the breakdown of adiabaticity in non-Hermitian systems if the eigenmode evolution tends to occur on the Riemann sheet with higher losses[15,31,34,40] (see Supplementary Fig. 11a,b). The features can also be related to the Stokes phenomenon of asymptotics[32,33]. The results of cases III and IV can be understood in the same way (see corresponding field distributions in Fig. 6g,h and eigenmode evolutions in Supplementary Fig. 11c-f) and demonstrate the chiral behavior in the encircling of one EP. In brief, the eigenmodes return to themselves in all four cases when zero or two EPs are encircled. When one EP is encircled, the output mode symmetry depends solely on the encircling direction (i.e., anti-symmetric output for counter-clockwise loops and symmetric output for clockwise loops), leading to a "chiral" behavior of the transmission. Note that the system is still reciprocal (i.e., $T_{nm} = T'_{mn}$) in the presence of the transverse bias field since the cross section of the coupled waveguides has a mirror symmetry. In addition, the field evolution in Fig. 6d,h indicates that the encircling is in fact not fully adiabatic. We show by simulation that a further increase in the system length can improve the results (Supplementary Fig. 12).

We perform microwave experiments to demonstrate the above effects. A photograph of the fabricated samples is shown in Fig. 7a (see the figure caption for detailed parameters). Only half of YIG waveguide 2 has the microwave absorber attached, which has been shown to be an effective way to minimize the dissipation for an encircling featuring the adiabatic transition while keeping the topology of the system intact[15] (Supplementary Fig. 13). We define a phase difference $\Delta\varphi = |\varphi_1 - \varphi_2|$ as the criterion to determine the symmetry of the output mode, where $\varphi_1$ ($\varphi_2$) is the phase measured at the output side of waveguide 1 (waveguide 2). For example, $\Delta\varphi = 0°$ corresponds to a symmetric mode whereas $\Delta\varphi = 180°$ an anti-symmetric mode. The measured phase differences as a function of the magnetic field strengths ($B_m$) and frequencies are shown in Fig. 7b,c, respectively, for cases I and III in which a symmetric mode is injected. We find in Fig. 7b that for each frequency there is a specific range of $B_m$ within which the system exhibits a state flip. This specific range shifts towards larger bias fields for lower frequencies. Figure 7d shows numerical simulation results for case I, which are highly consistent with the measurement. We also calculate for each frequency the positions of the EPs and mark them with the two white dashed lines in Fig. 7d. The whole map can be partitioned into three regions depending on the number of EPs encircled, confirming that the state flip indeed stems from the encircling of one EP. In contrast, the output in case III is always a symmetric mode regardless of the number of EPs encircled, which is hence an experimental demonstration of the breakdown of adiabaticity when encircling one EP (see Fig. 7c for experimental results and Fig. 7e for simulation results). The phase differences for cases II and IV injected with an anti-symmetric mode are shown in Fig. 8, which, together with the results in Fig. 7, demonstrate the behavior of mode switching when different numbers of EPs are encircled. Results of a control experiment for dynamical encircling are given in Supplementary Fig. 14.

The chiral nature of the dynamics of encircling one EP has been exploited for asymmetric mode switching[15]. Since the external field in this work can be tuned continuously, our system can be applied to the switching of modes controlled with external fields, i.e., manipulating the symmetry of the output state by dynamically encircling different numbers of EPs. Our system is a good platform for exploring non-Hermitian physics since it is the first design that allows for the experimental realization of a dynamical encircling of two EPs. Our numerical and experimental results convincingly demonstrate that the encircling of two EPs for all the four cases is adiabatic in the presence of non-Hermiticity, ensuring that the eigenstate returns to itself after looping around the two EPs that are spawn from a single DP.

## Discussion

We have shown both theoretically and experimentally that in a two-state coupled ferromagnetic waveguide system, tuning an external magnetic field creates an EP and a subsequent symmetry recovery, the latter of which is rarely seen in two-state systems. Based on the unique topological structure of the proposed system possessing a pair of EPs, we have introduced a parameter space in which the EPs can be dynamically encircled. The measured mode symmetry under different bias fields serves as direct evidence for the presence of the EP and symmetry recovery. We emphasize that the bias field can be used in situ to continuously tune the size of the loop, i.e., from excluding any EP to dynamically encircling one or even two EPs. The proposed system can thus be applied to mode switching controlled with an external parameter without changing or moving the sample.

## Methods

**Sample preparation.** In the experiment demonstrating the presence of an EP and symmetry recovery (Fig. 4), each waveguide consisted of four YIG samples with the dimensions of 8 mm × 4 mm × 50 mm and used as purchased. In the experiment demonstrating the encircling of EP and symmetry recovery (Figs 7,8), YIG waveguide 1 consisted of eight YIG samples with the dimensions of 8 mm × 4 mm × 50 mm. YIG waveguide 2 also consisted of eight YIG samples, but each one was polished from a larger sample (16 mm × 4 mm × 50 mm) using a hand polishing machine to the shape shown in Fig. 7a (also see Fig. 5d).

**Experimental setup.** In the experiment demonstrating the presence of an EP and symmetry recovery (Fig. 4), we placed an antenna ~8 mm in length near the surface of the lossy waveguide in order to excite both eigenmodes in the system. Another antenna ~20 mm in length was placed at the waveguide exit to receive the transmission. The two antennas were connected to an Agilent Technologies 8720ES Network Analyzer for recording the transmission intensity. In the experiment demonstrating the encircling of EP and symmetry recovery (Figs 7,8), the symmetric mode was excited using an ~20 mm long antenna, while the anti-symmetric mode was excited using two ~8 mm long antennas. These two antennas were connected to the coaxial cable of the 8720ES Network Analyzer via a one-to-two power splitter and were placed along opposite directions so that their currents were oscillating out of phase. An antenna ~8 mm in length was placed at the exit of waveguide 1 and that of waveguide 2 to detect their corresponding phases $\varphi_1$ and $\varphi_2$. The Agilent 85071C material measurement software was used to measure the permittivity of the microwave absorber (Supplementary Fig. 4b). The VSM was used to measure the hysteresis loop of the YIG sample (Supplementary Fig. 4a).

**Numerical simulations.** In the numerical simulations of Figs 1,2, the permeability tensor of the ferromagnetic material takes the form

$$\mu_1 = \begin{bmatrix} \mu_b & 0 & i\chi \\ 0 & 1 & 0 \\ -i\chi & 0 & \mu_b \end{bmatrix}, \ \mu_2 = \begin{bmatrix} \mu_b & 0 & -i\chi \\ 0 & 1 & 0 \\ i\chi & 0 & \mu_b \end{bmatrix}. (1)$$

All simulations were performed using the commercial solver package COMSOL. The effective mode index and eigenmode profiles (e.g., Figs 1b,2b) were calculated using the mode analysis study of the RF Module. The transmission spectra and field distributions (e.g., Fig. 3c,d) were simulated using the frequency domain study of the RF Module. The transmission intensities given by $T_{nm} = |t_{nm}|^2$ (e.g., Fig. 6b) were calculated using the frequency domain study associated with the boundary mode analysis of the RF Module. In the simulations of Figs 7d,e and 8c,d, the phase difference was calculated based on the obtained transmission intensities (e.g., $\Delta\varphi = 2\arctan\left(T_{21}/T_{11}\right)$ for case I), while the permittivity of the microwave absorber was simply chosen as 3+10*i* for all frequencies given that the loss is smaller than the one used in Fig. 4. There was actually a Fabry-Perot effect in the experiment which was however not considered in the simulations of Figs. 6b-h,7d,e,8c,d. We show in Supplementary Fig. 15 that the effects arising from the encircling of EPs would not be disturbed by the Fabry-Perot effect seen in the experiment.

## ACKNOWLEDGEMENTS

This work was supported by the Hong Kong Research Grants Council through grant AoE/P-02/12. Xu-Lin Zhang is also supported by the National Natural Science Foundation of China (grant no. 61605056) and the China Postdoctoral Science Foundation (grant no. 2016M591480).

## Figures and captions

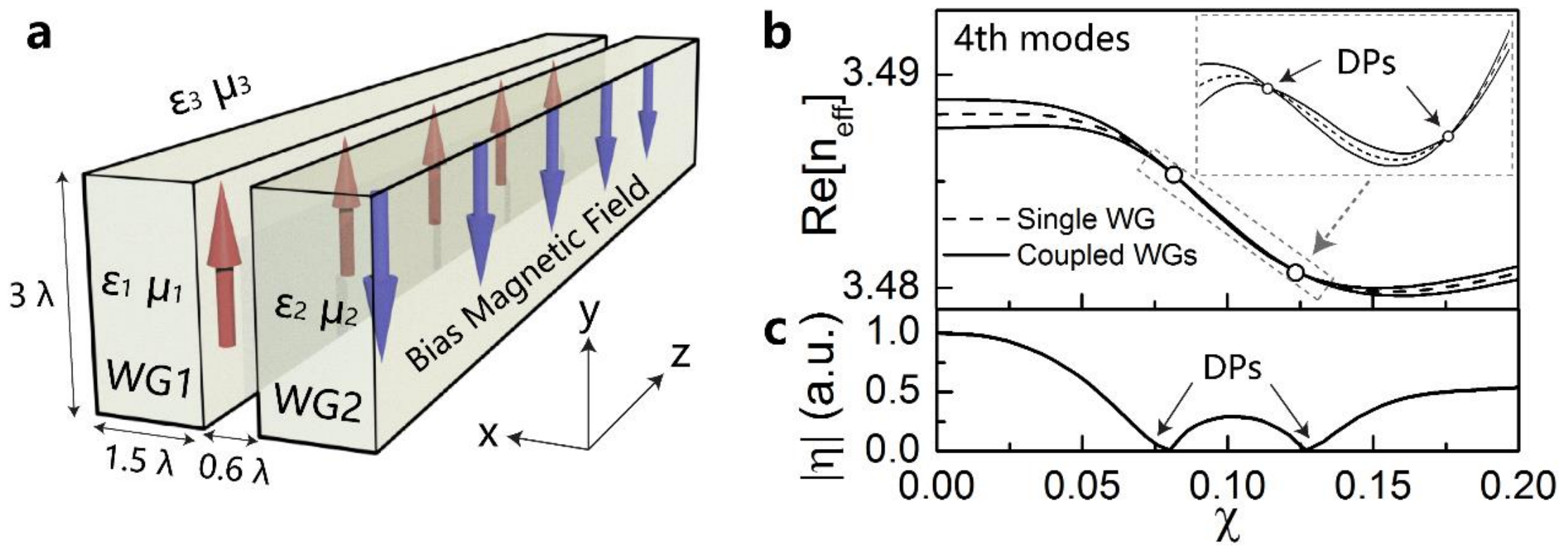

**Figure 1 | Degeneracy points in lossless coupled ferromagnetic waveguides. (a)** Schematic diagram of a coupled ferromagnetic waveguide system with bias magnetic fields applied along opposite directions. **(b)** Calculated effective mode index of the fourth pair of modes in the coupled system (solid lines) and that of the fourth mode in a single waveguide (dashed line) as a function of $\chi$ with $\gamma = 0$, where two DPs can be found. **(c)** Calculated mode coupling coefficient $|\eta|$ (see text for definition).

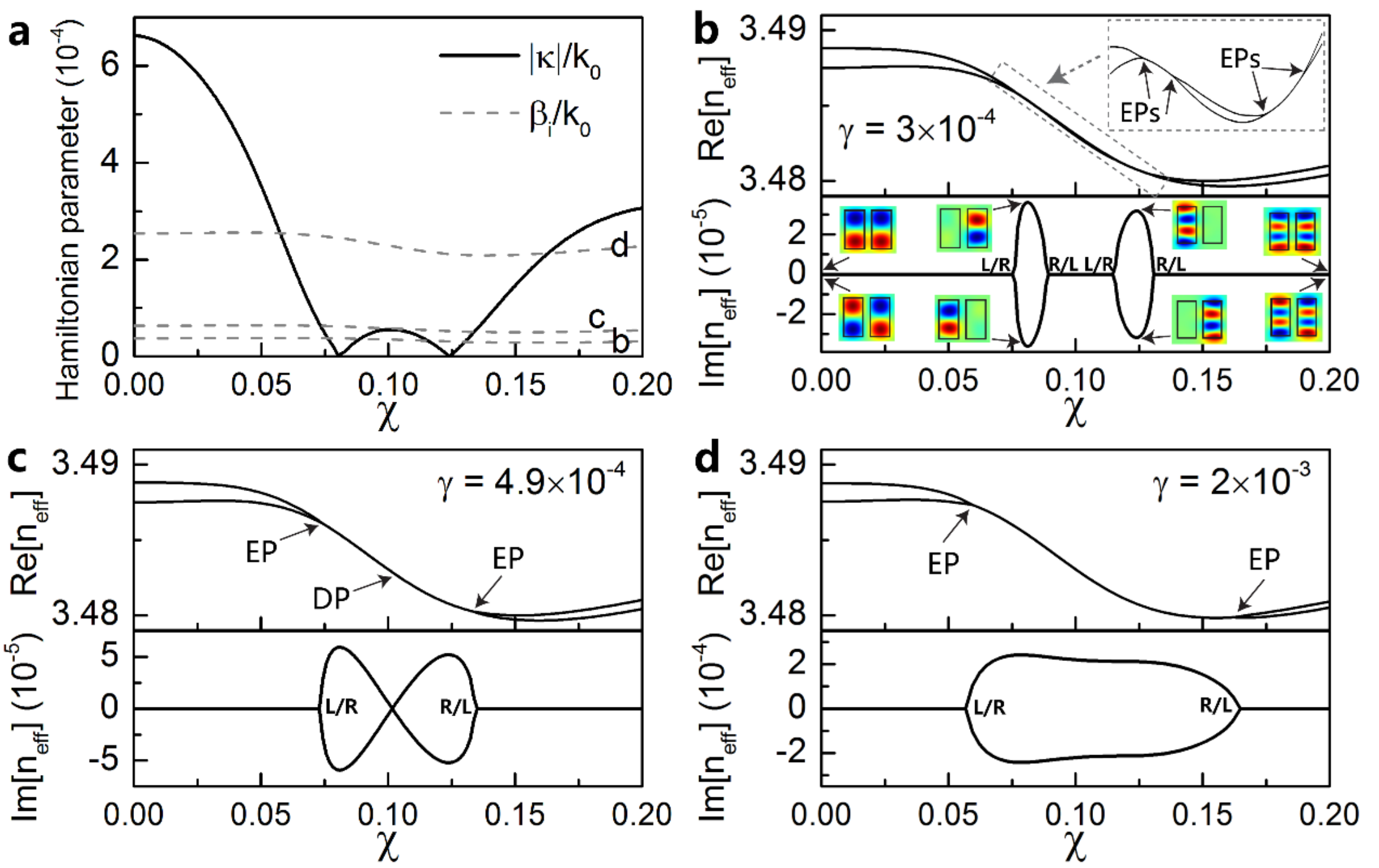

**Figure 2 | Exceptional points in PT-symmetric coupled ferromagnetic waveguides.** **(a)** The coupling parameter $\kappa$ (solid line) obtained through fitting for the fourth pair of modes in the system without gain or loss. The dashed lines show the calculated imaginary part of the effective mode index of the fourth mode in a single waveguide with gain and loss corresponding to those in **b-d**. **(b-d)** Calculated effective mode index of the fourth pair of modes with different gain and loss, **(b)** $\gamma = 3\times10^{-4}$, **(c)** $\gamma = 4.9\times10^{-4}$, and **(d)** $\gamma = 2\times10^{-3}$. We find multiple EPs and symmetry recovery behaviors. The inset of **b** shows $H_y$ eigenfield distributions at different $\chi$, showing features typical of *PT*-symmetric and *PT*-broken regions.

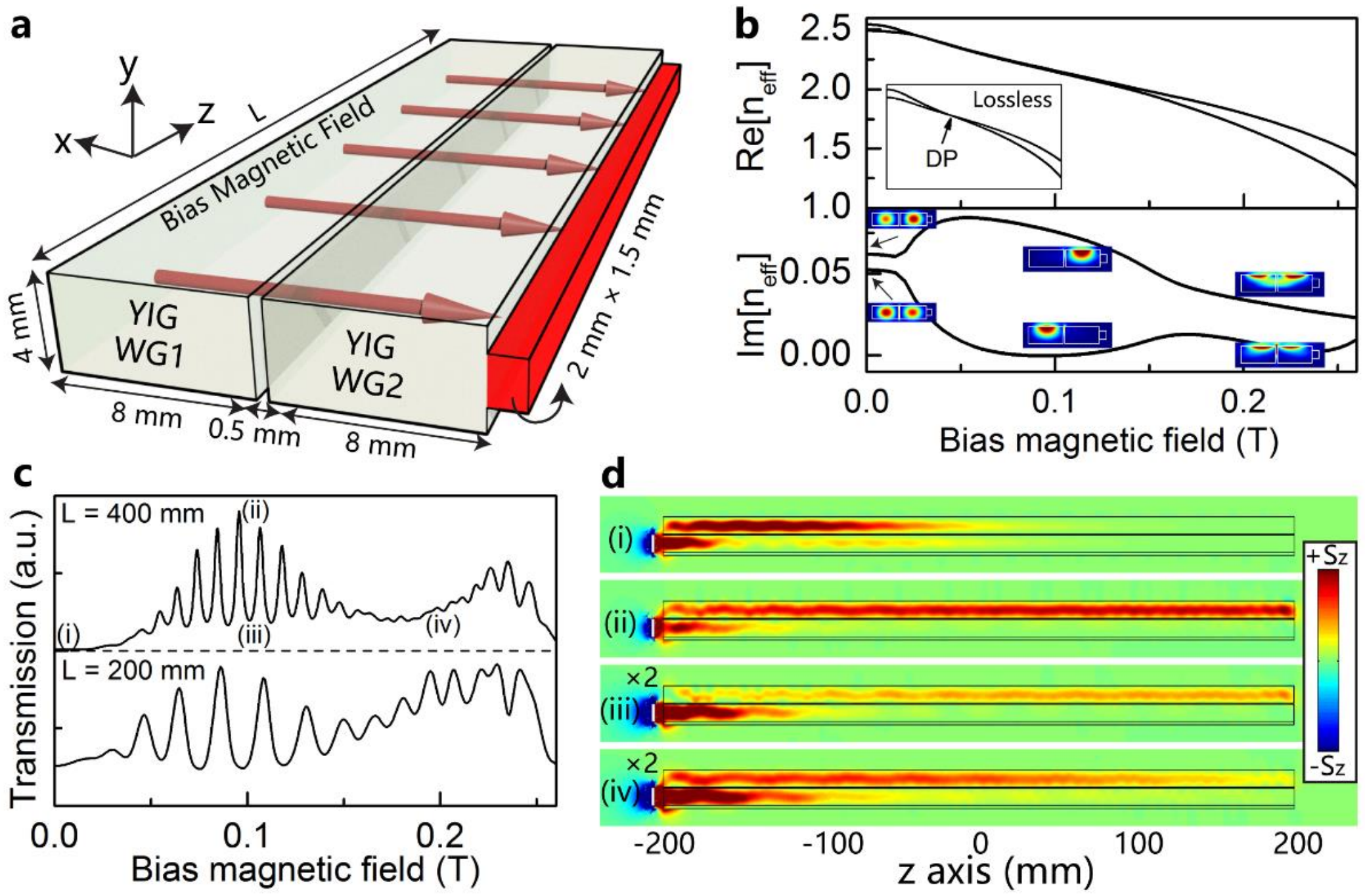

**Figure 3 | Numerical demonstration of an exceptional point and symmetry recovery.** **(a)** Schematic diagram of a passive coupled YIG waveguide system with microwave absorbers attached to the side of one waveguide. **(b)** Calculated effective mode index as a function of the bias magnetic field, in which we find the salient feature of an EP as well as symmetry recovery. The inset of the upper panel shows the results for the lossless coupled system, where a DP appears and can spawn a pair of EPs when an absorber is added. The six colored patterns in the lower panel show the power flow distributions in the *x*-*y* plane under different bias magnetic fields. **(c)** Numerically simulated transmission spectra as a function of the magnetic field for *L* = 200 mm and 400 mm, where a region with enhanced transmission can be found that matches well with the broken phase region predicted in **b**. The peaks correspond to Fabry-Perot resonances. **(d)** Power flow distributions in the coupled system with *L* = 400 mm for different magnetic fields marked as (i)-(iv) in **c**. The field patterns in (iii) and (iv) have been scaled up by a factor of two for improved readability. The frequency is 9.5 GHz in all of the simulations.

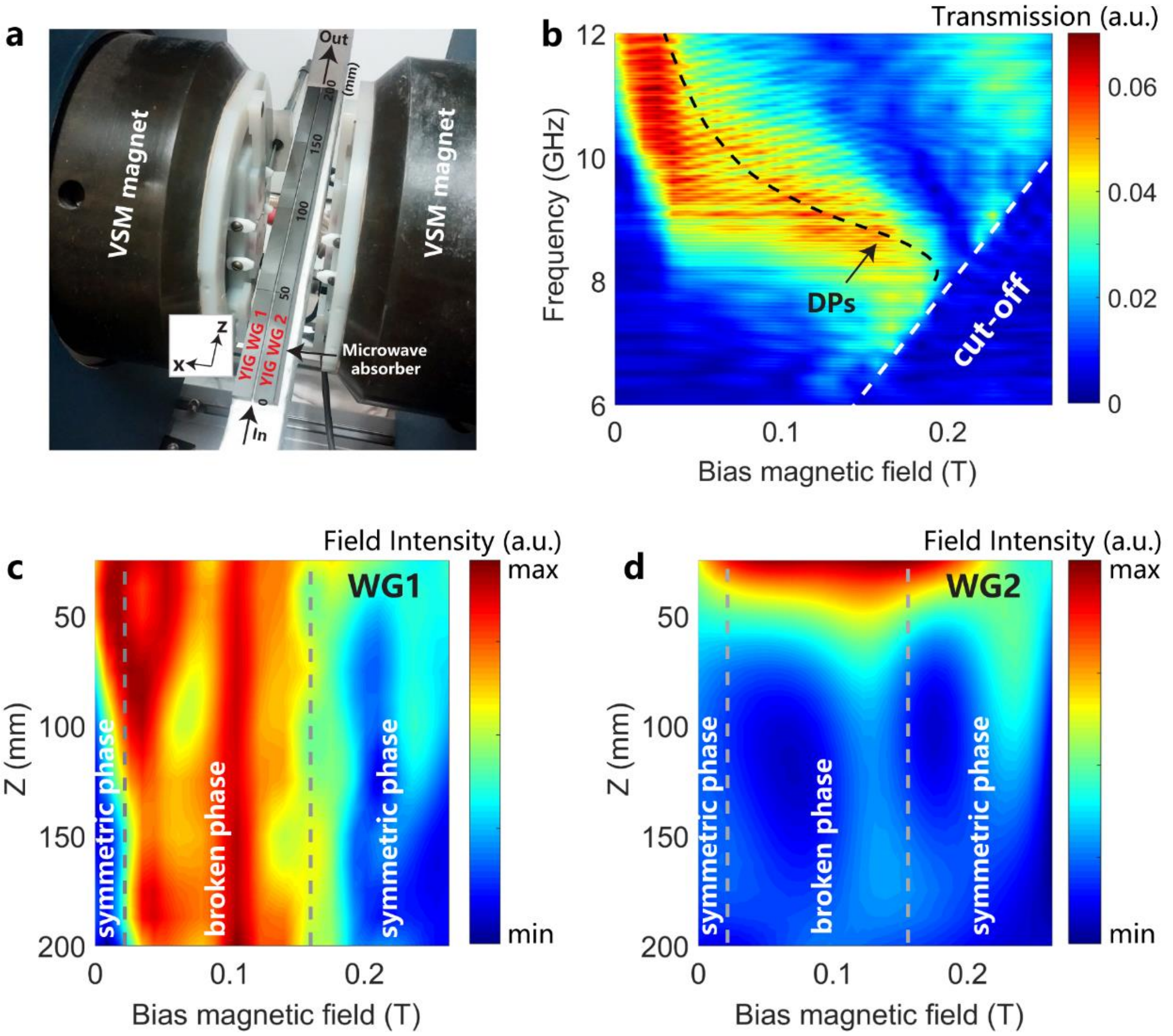

**Figure 4 | Experimental demonstration of the exceptional point and symmetry recovery. (a)** A photograph of the coupled YIG waveguide system placed inside the VSM. Microwave absorbers are attached to the side of one waveguide. Each waveguide measures 8 mm × 4 mm × 200 mm and the gap distance is 0.5 mm. **(b)** Experimentally measured transmission spectra for different magnetic field strengths and frequencies, where the black dashed line shows the calculated DPs. The modes are cut off in the region below the white dashed line. **(c,d)** Experimentally measured surface electric field intensity along the propagation direction (see **a** for the definition of *z*-axis) at 9.5 GHz for **(c)** waveguide 1 and **(d)** waveguide 2. The two grey dashed lines mark the boundaries between symmetric phase and broken phase regimes.

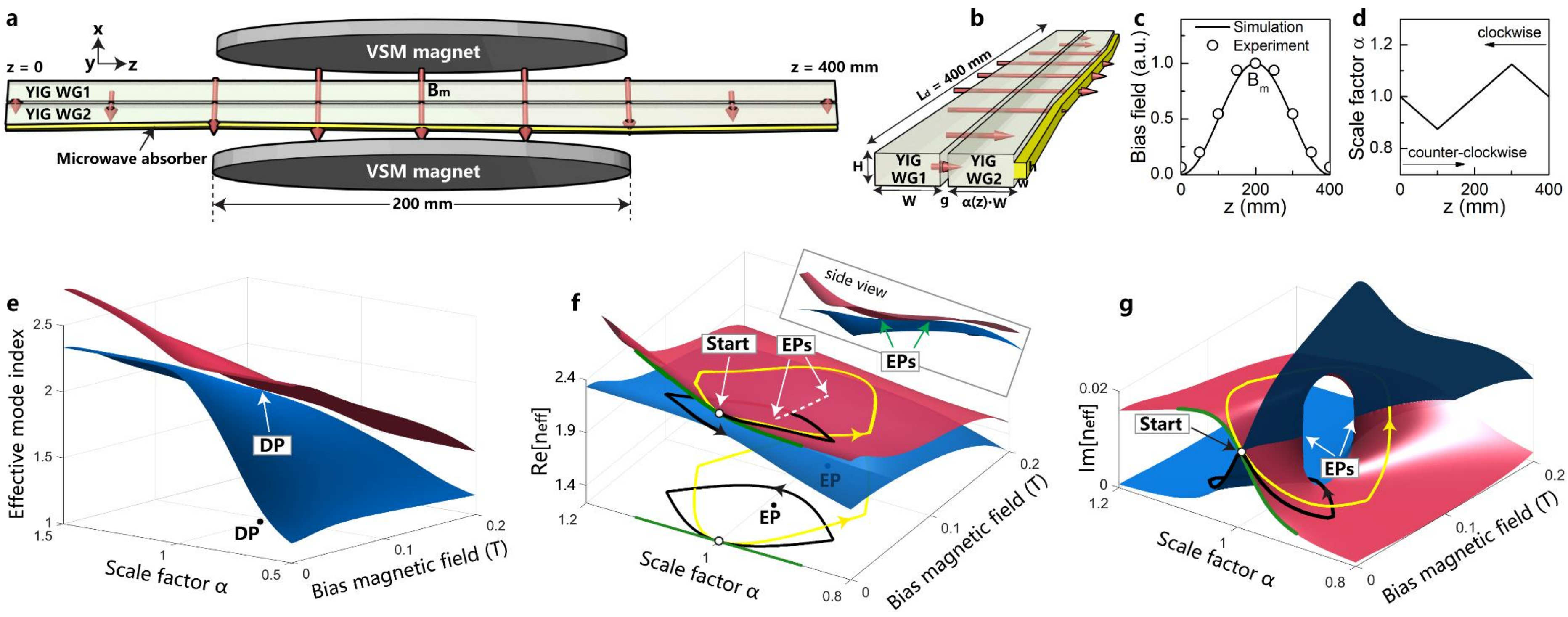

**Figure 5 | Design of system for dynamically encircling EPs and corresponding Riemann surface. (a)** Schematic diagram of a coupled ferromagnetic waveguide system with length $L_d$ = 400 mm, where the bias field generated by the two VSM magnets and the width of YIG waveguide 2 vary continuously along the $z$-axis. **(b)** Side view of the coupled system with structural parameters $W$ = 8 mm, $H$ = 4 mm, $g$ = 1 mm, $w$ = 1.5 mm, and $h$ = 2 mm. **(c)** Experimentally measured bias field distributions along the $z$-axis (circles), fitted using $B(z) = B_m \sin\left(\pi z / L_d\right)$ for numerical simulations (solid line). **(d)** Variation in the scale factor $\alpha$ along the $z$-axis. The minimum $\alpha$ is 0.875 (at $z$ = 100 mm) and the maximum

is 1.125 (at $z$ = 300 mm), corresponding to the widths of 7 mm and 9 mm, respectively, for YIG waveguide 2. **(e)** Calculated effective mode index as a function of the bias field and scale factor at 9 GHz for the system without the microwave absorber. A DP appears ($B$ = 0.092 T, $\alpha$ = 1) due to the absence of mode couplings. **(f,g)** Calculated real part **(f)** and imaginary part **(g)** of the effective mode index as a function of the bias field and scale factor at 9 GHz for the lossy system. The two figures show a self-intersecting Riemann surface with the EP located at $B$ = 0.06 T, $\alpha$ = 0.988 and symmetry recovery located at $B$ = 0.123 T, $\alpha$ = 0.982. The permittivity of the absorber is chosen as 3+3$i$. The white dashed line in **f** marks the 'exact' broken phase region where the real parts of the two eigenvalues coalesce. The green, black, and yellow lines represent counter-clockwise loops (starting with a symmetric mode at $B$ = 0, $\alpha$ = 1) generated by $B_m$ = 0 T, 0.08 T, and 0.17 T, respectively. A flip occurs between the two eigenstates for the black loop which encircles one EP. The eigenstate returns to itself for the green and yellow loops where zero and two EPs are encircled, respectively.

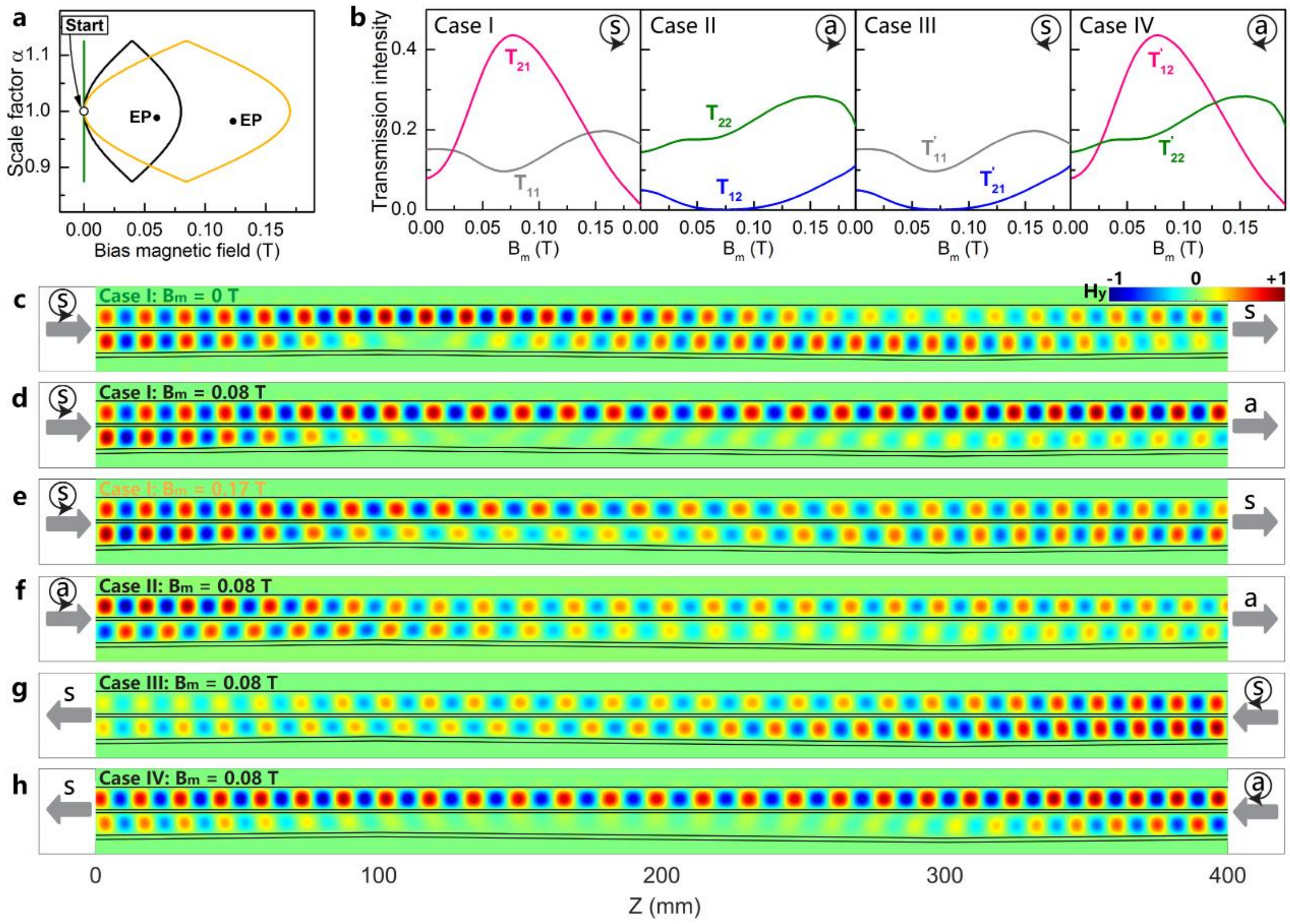

**Figure 6 | Numerical demonstration of the dynamical encircling of EPs. (a)** Parameter space in which three loops are generated by $B_m$ = 0 T (green loop), 0.08 T (black loop), and 0.17 T (yellow loop), with the starting/end point at $B$ = 0 and $\alpha$ = 1. Injections from $z$ = 0 and $z$ = 400 mm correspond to counter-clockwise and clockwise loops, respectively. **(b)** Calculated transmission intensities for the four cases (see text for definition) as a function of $B_m$. The state flip can only occur for cases I and IV when one EP is encircled, otherwise the symmetry of the eigenmode at the waveguide exit is the same as that of injection. **(c-h)** Numerically simulated $H_y$ field distributions for different input modes and injection directions. The results for case I are shown in **c-e** with $B_m$ = 0 T, 0.08 T, and 0.17 T, respectively. Shown in **f-h** are results for cases II, III, and IV, respectively, with $B_m$ = 0.08 T. The frequency is 9 GHz in all of the simulations.

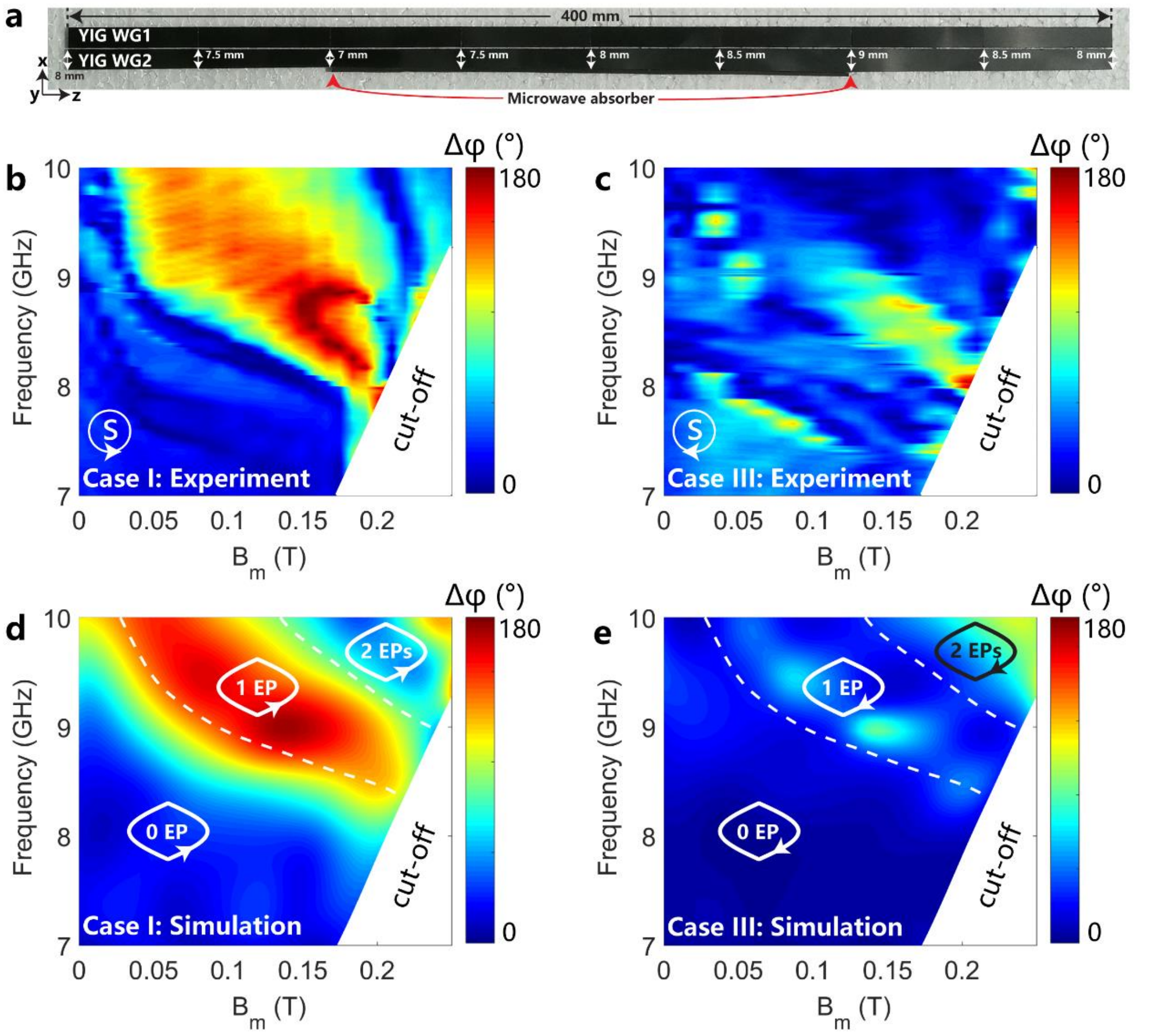

**Figure 7 | Experimental demonstration of the dynamical encircling of EPs when a symmetric mode is injected. (a)** A photograph of the fabricated coupled YIG waveguides. YIG waveguide 1 measures $W \times H \times L_d$ = 8 mm × 4 mm × 400 mm, while the dimensions of YIG waveguide 2 are $\alpha(z)W \times H \times L_d$ with the profile of $\alpha(z)$ shown in Fig. 5d. The gap distance is ~0.5 mm. Microwave absorbers with the dimensions of ~2 mm × 1 mm × 200 mm are attached to the side of YIG waveguide 2 to introduce asymmetric losses. **(b,c)** Experimentally measured phase differences $\Delta\varphi$ for various bias fields $B_m$ and frequencies for case I **(b)** and case III **(c)**. **(d,e)** Numerically simulated phase differences $\Delta\varphi$ as a function of the bias fields $B_m$ and frequencies for case I **(d)** and case III **(e)**. The two dashed lines mark the calculated positions of EPs which partition the map into three regions depending on the number of EPs encircled. The phase differences in the region where the eigenmodes are cut off are not shown since the results there are meaningless.

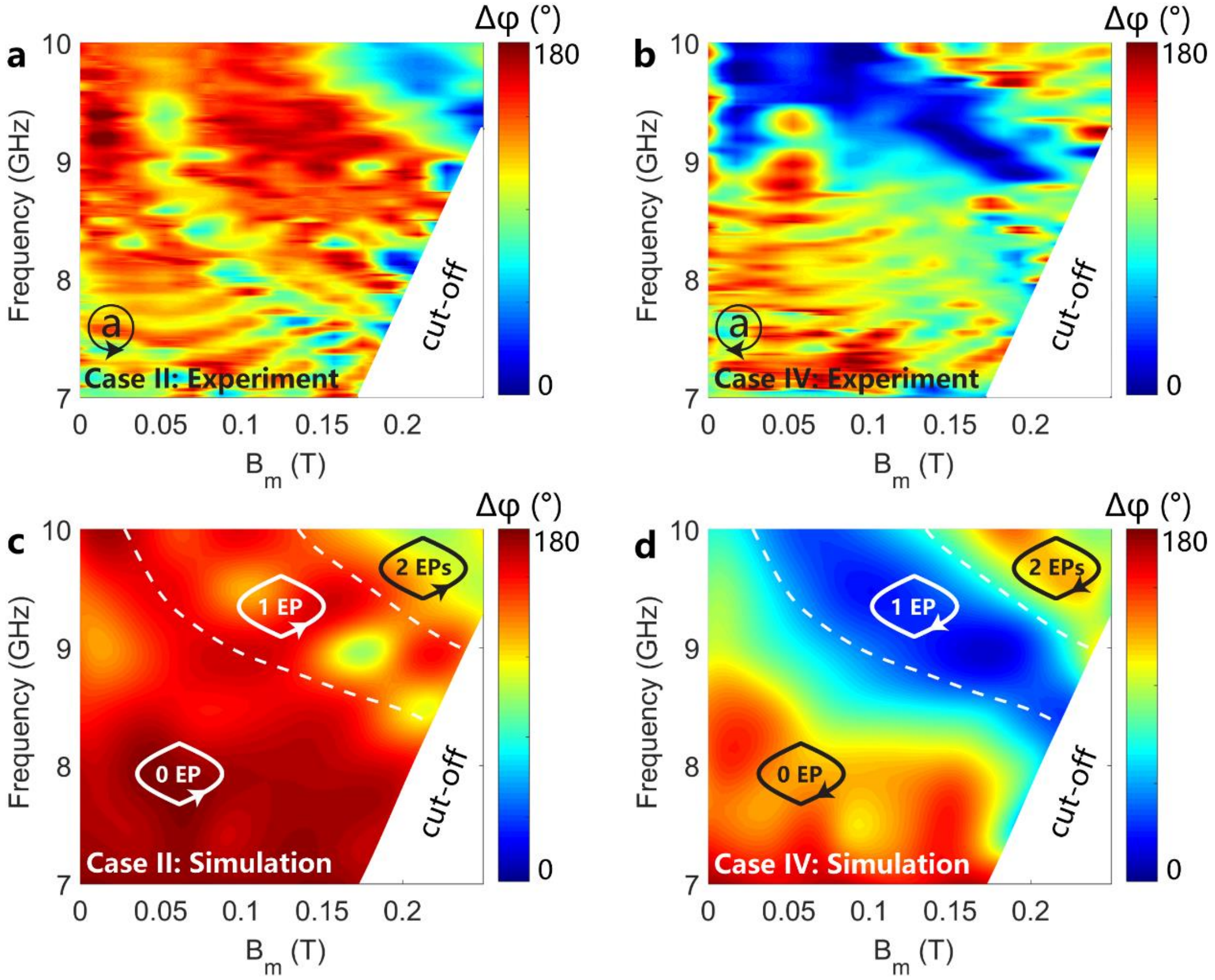

**Figure 8 | Experimental demonstration of the dynamical encircling of EPs when an anti-symmetric mode is injected. (a,b)** Experimentally measured phase differences $\Delta\varphi$ for various bias fields $B_m$ and frequencies for case II **(a)** and case IV **(b)**. **(c,d)** Numerically simulated phase differences $\Delta\varphi$ as a function of the bias fields $B_m$ and frequencies for case II **(c)** and case IV **(d)**.

## Supplementary Figures

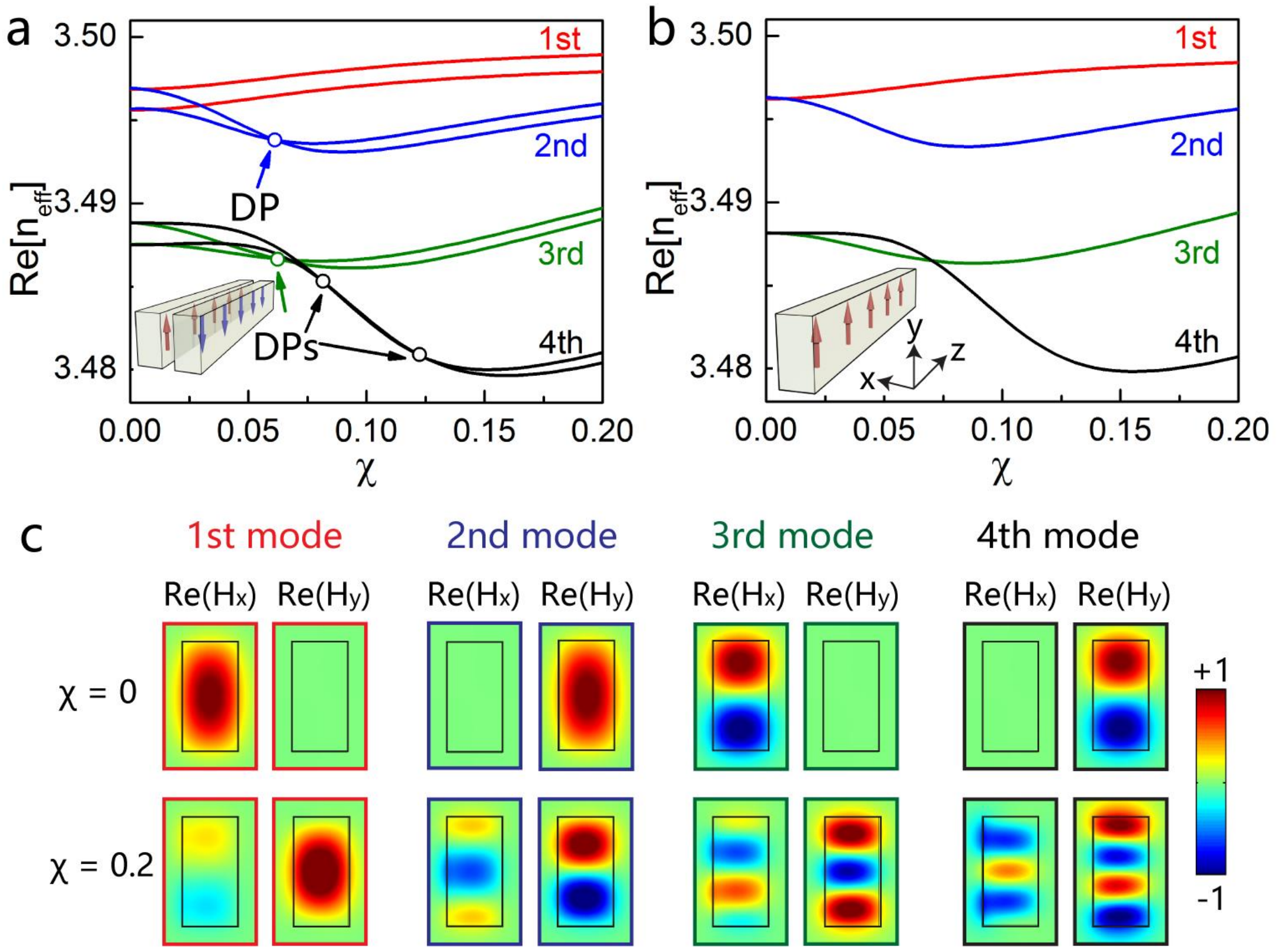

**Supplementary Figure 1 | Eigenvalue dispersions and eigenfield distributions in the lossless system with opposite directions of bias fields. (a)** Calculated effective mode index of the first four pairs of modes in the coupled waveguides as a function of $\chi$. **(b)** Calculated effective mode index of the first four modes in the single waveguide as a function of $\chi$. **(c)** Magnetic field distributions for the first four modes in the single waveguide with $\chi$ = 0 and 0.2.

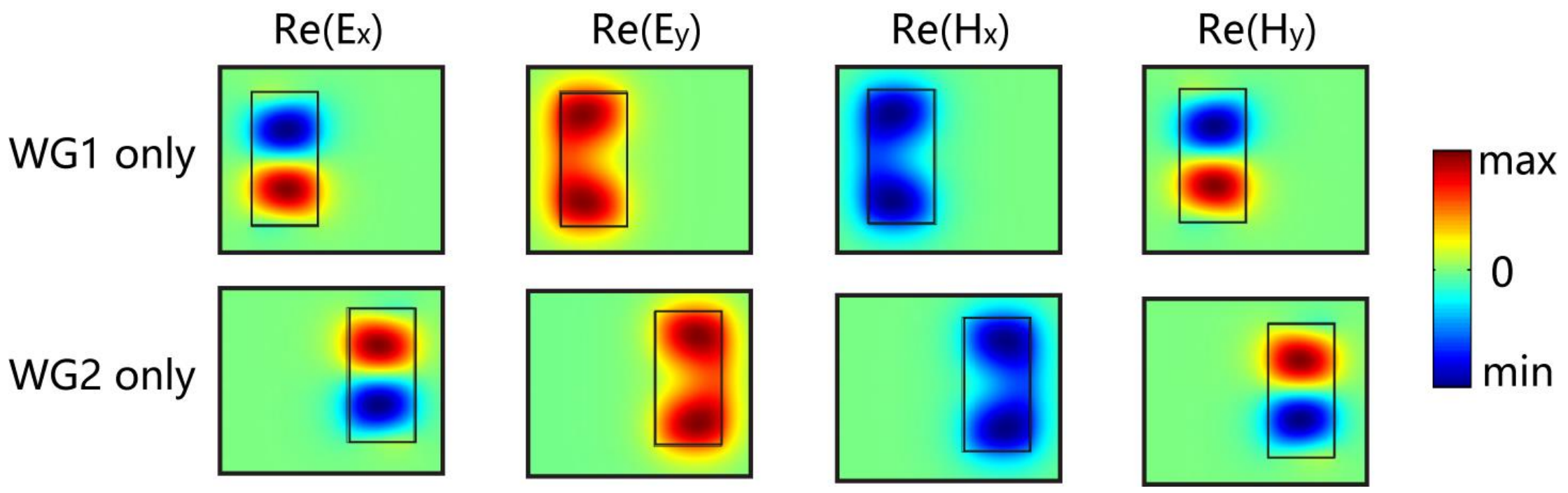

**Supplementary Figure 2 | Eigenfield distributions of the uncoupled mode in the lossless system.** Eigenfield patterns of the two uncoupled fourth set of modes in a single waveguide with $\chi$ = 0.0816. At the same value a DP appears between the fourth pair of modes in the coupled waveguides. The integration of the field patterns gives a zero-coupling strength. So, if we use the eigenmodes of a single waveguide to predict the DP of coupled waveguides, in the spirit of standard perturbation theory, we will predict a value of $\chi$ = 0.0816.

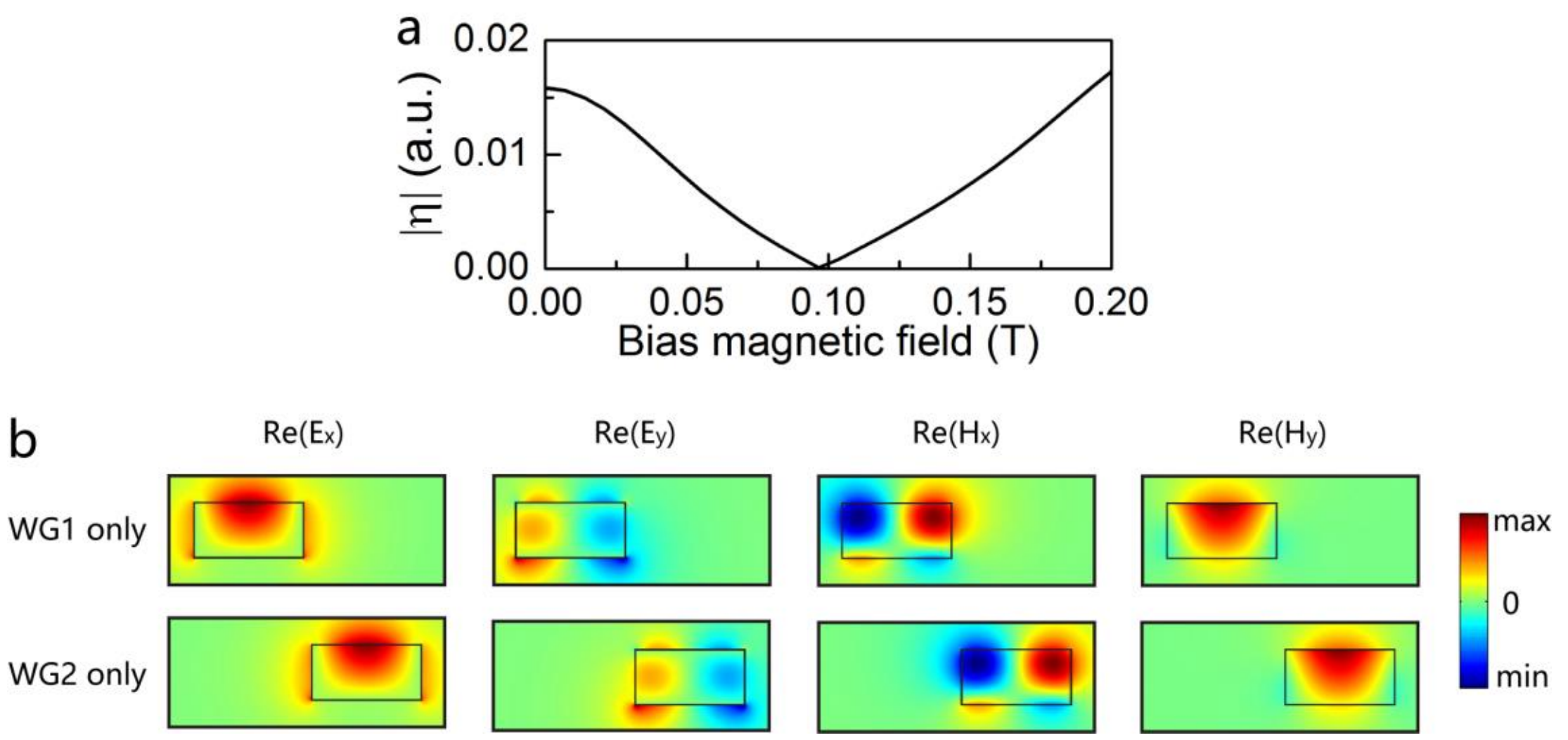

**Supplementary Figure 3 | Eigenfield distributions at the degeneracy point in the lossless system. (a)** Coupling coefficient between the two uncoupled fundamental modes in the coupled system shown in Fig. 3a of the main text. The frequency is 9.5 GHz. **(b)** Eigenfield patterns of the two uncoupled modes in a single YIG waveguide with a bias magnetic field of ~0.095 T, corresponding to the case of the DP in **a**.

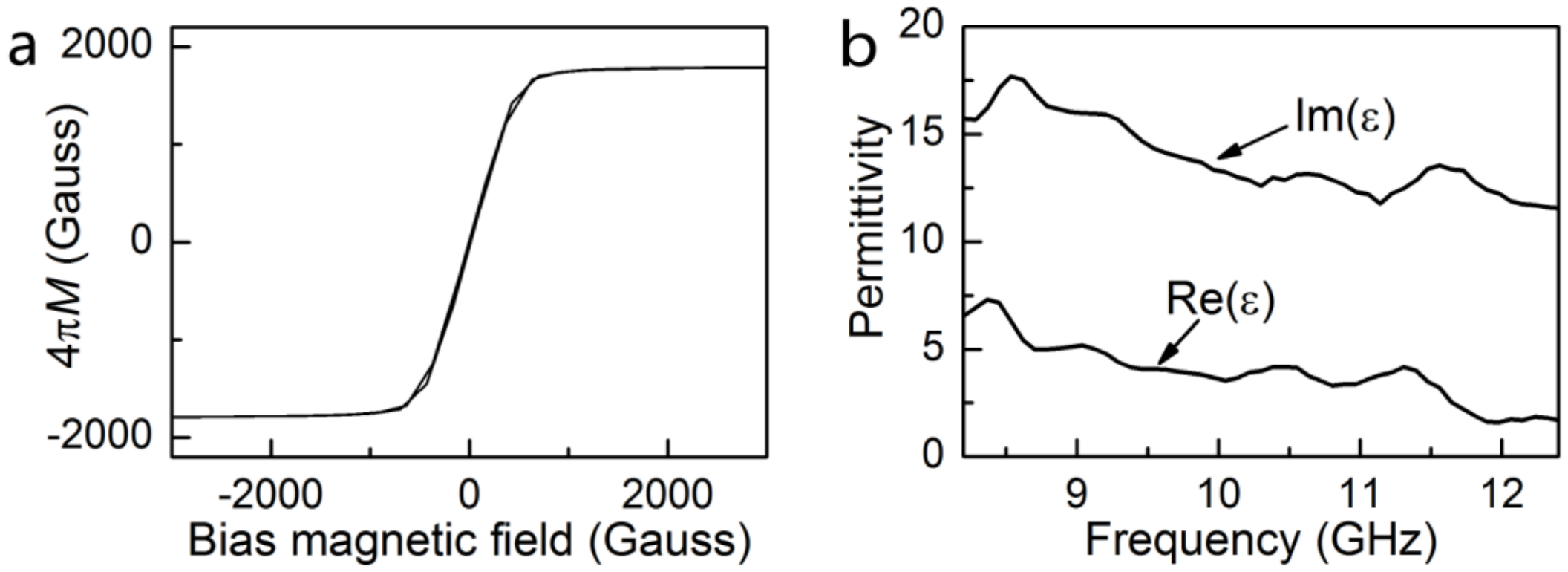

**Supplementary Figure 4 | Materials properties of YIG and the microwave absorbers.** **(a)** Measured hysteresis loop of YIG using the VSM, where $M$ is the measured magnetization. This curve is then used for calculating the permeability tensor of YIG. **(b)** Measured permittivity of the microwave absorber. In fact, we have a series of microwave absorbers exhibiting different absorption coefficients. This figure shows the permittivity of the absorber with the largest absorption coefficient, which was used for the experiment in Fig. 4 of the main text. The absorber used for the experiment in Figs. 7,8 of the main text has a smaller absorption coefficient than this one so that in the corresponding simulations we simply used $\varepsilon = 3+10i$ for all the frequencies (also see Methods).

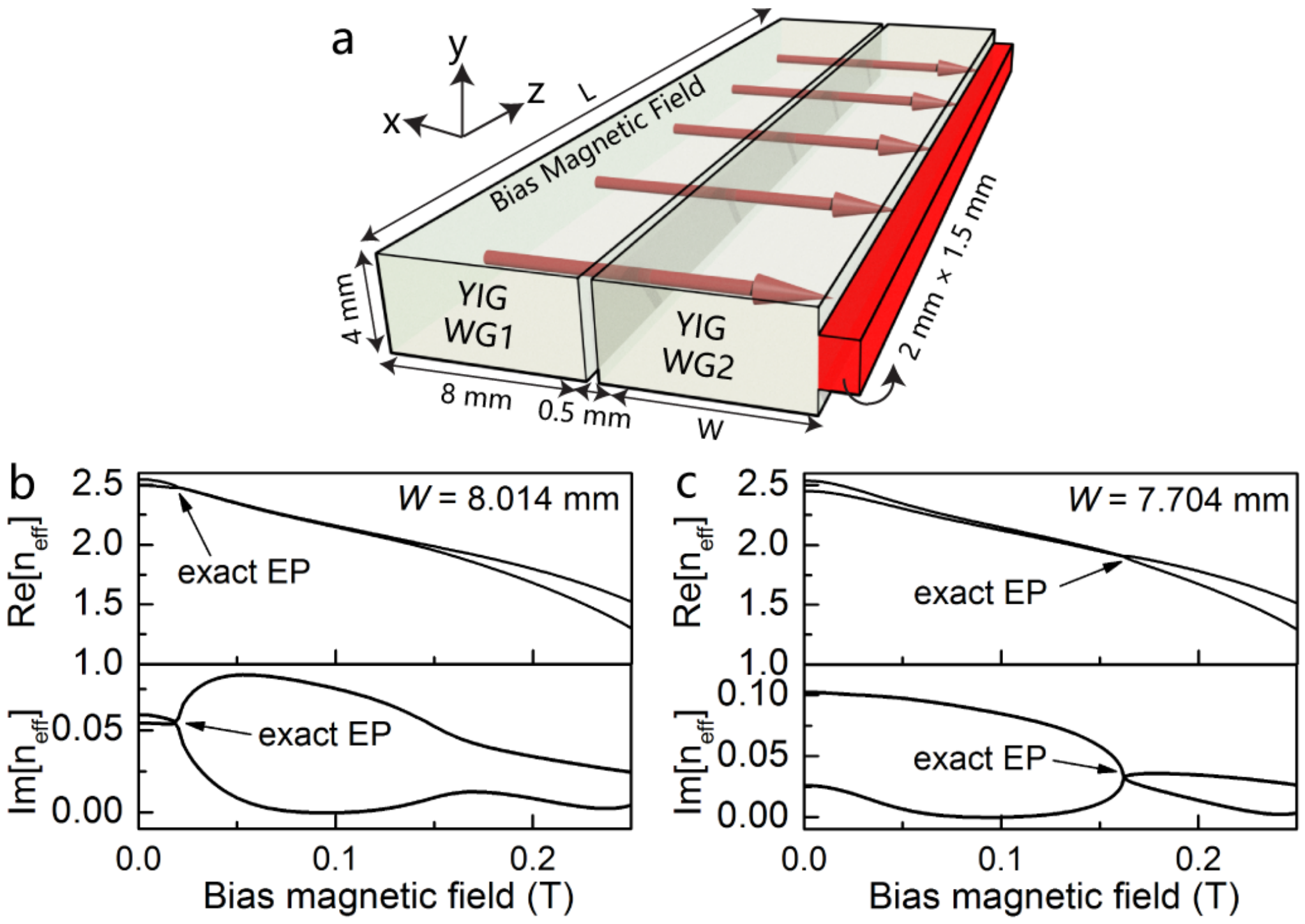

**Supplementary Figure 5 | 'Exact' exceptional point and symmetry recovery point. (a)** Schematic diagram of a YIG coupled waveguide system in which we tune the width of waveguide 2 in order to reach the 'exact' EPs. **(b)** Calculated effective mode index as a function of the bias magnetic field with $W$ = 8.014 mm where the 'exact' EP can be found at 0.019 T. **(c)** Calculated effective mode index as a function of the bias magnetic field with $W$ = 7.704 mm where the 'exact' symmetry recovery point can be found at 0.162 T. Although it is impractical to realize 'exact' EPs in experiments (due to limitations such as fabrication precision), creating a system that is very close to the 'exact' EPs as we show in this work is sufficient to study the physics and consequences of EPs. These results are in fact the basis for creating a parameter space in which the EPs are encircled. We note the EPs in Fig. 6a of the main text are also slightly away from the 'real axis' (i.e., $\alpha$ = 1).

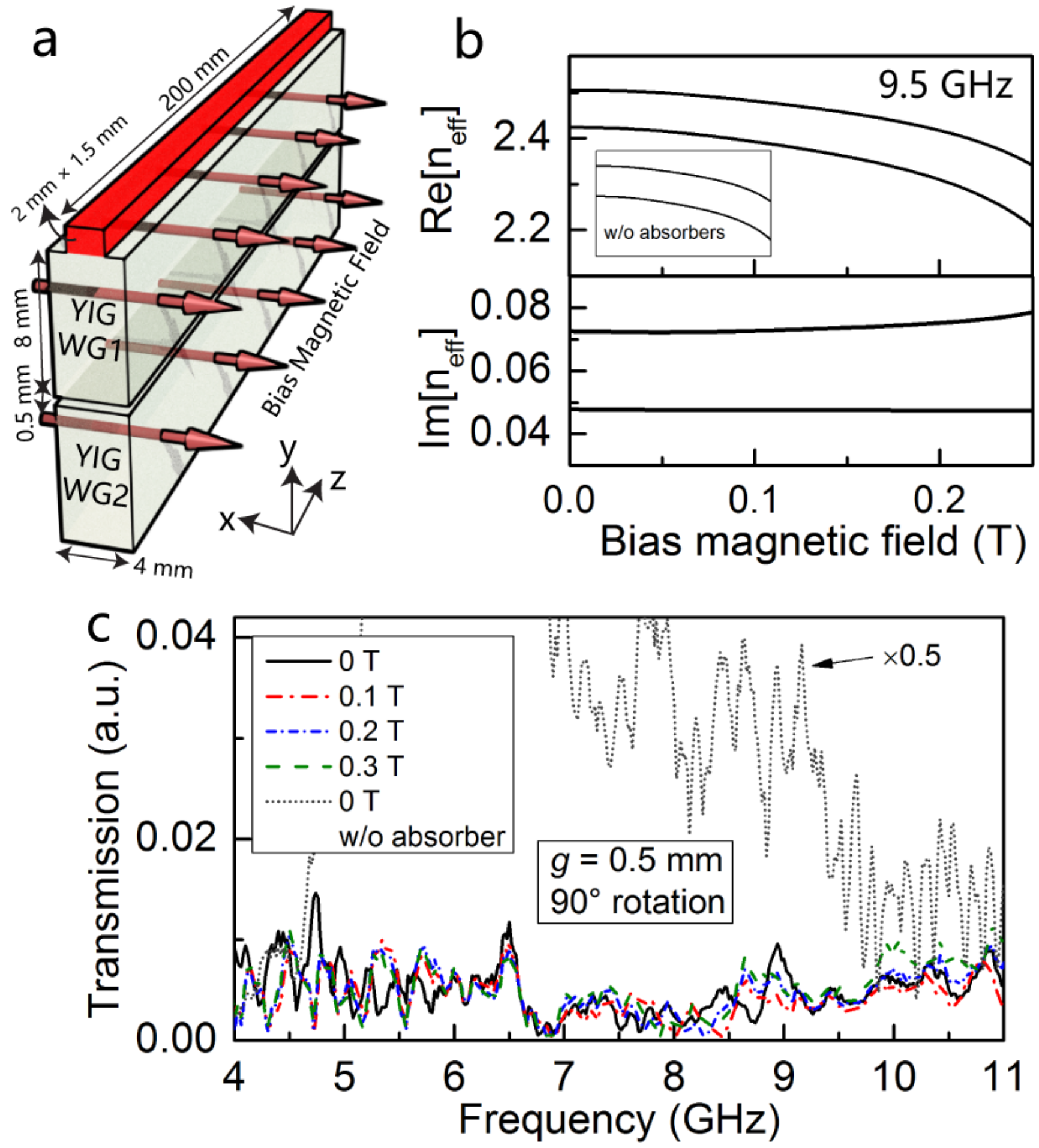

**Supplementary Figure 6 | Control experiment I. (a)** Schematic diagram of a coupled ferromagnetic waveguide system with the bias magnetic field parallel to the short-edge of the waveguide. **(b)** Calculated effective mode index as a function of the bias magnetic field. The inset shows the real part of the effective mode index in the lossless system, showing the absence of DPs. **(c)** Measured transmission spectra with different bias fields, where the dashed line shows the result for the system without the absorber.

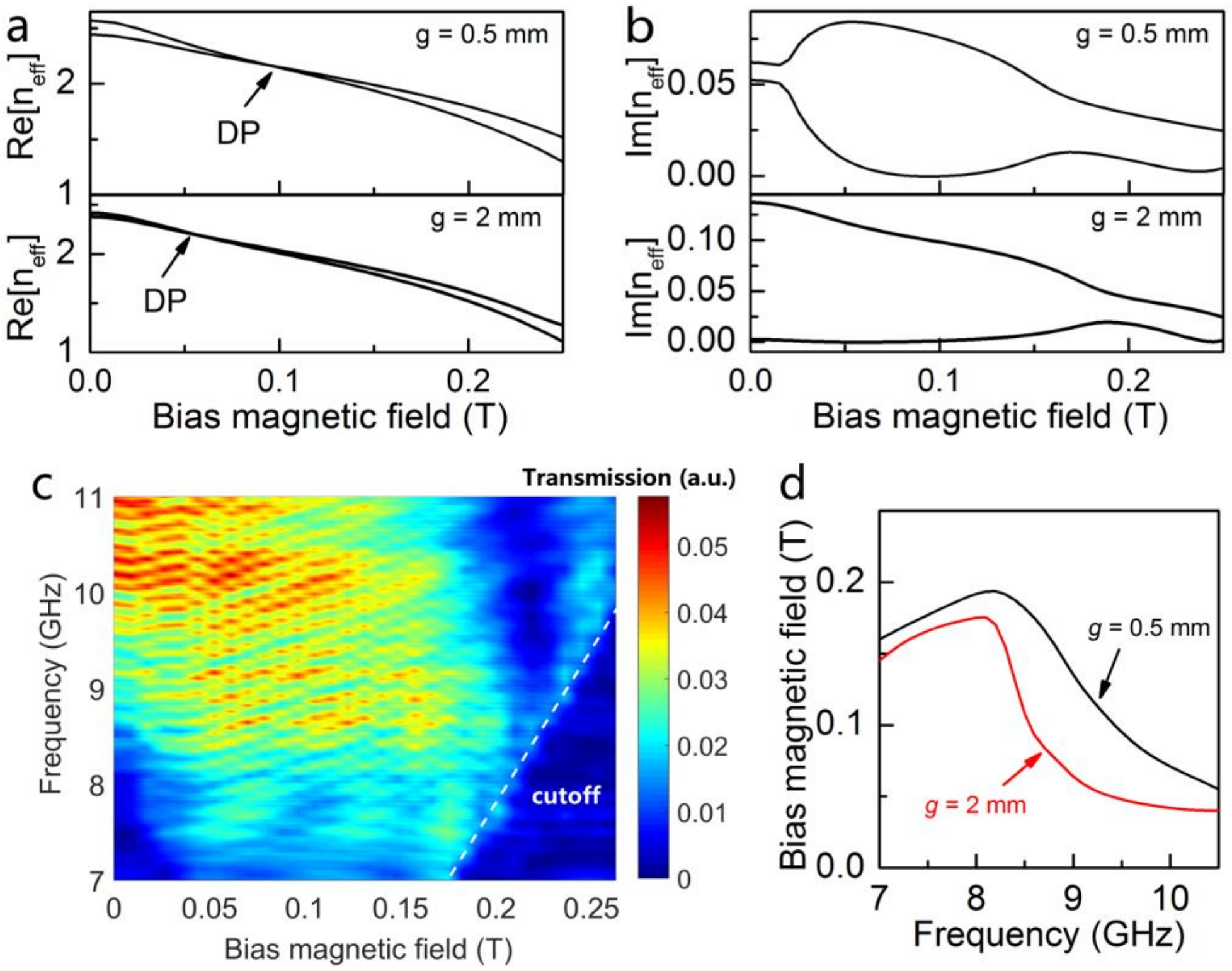

**Supplementary Figure 7 | Control experiment II.** Calculated effective mode index as a function of the bias magnetic field in the lossless system **(a)** and lossy system **(b)** with different gap distances at 9.5 GHz. **(c)** Measured transmission spectra of the 2 mm-gap system for different bias fields and frequencies. **(d)** Calculated bias magnetic field required to reach the DP in the lossless system with *g* = 0.5 mm and *g* = 2 mm.

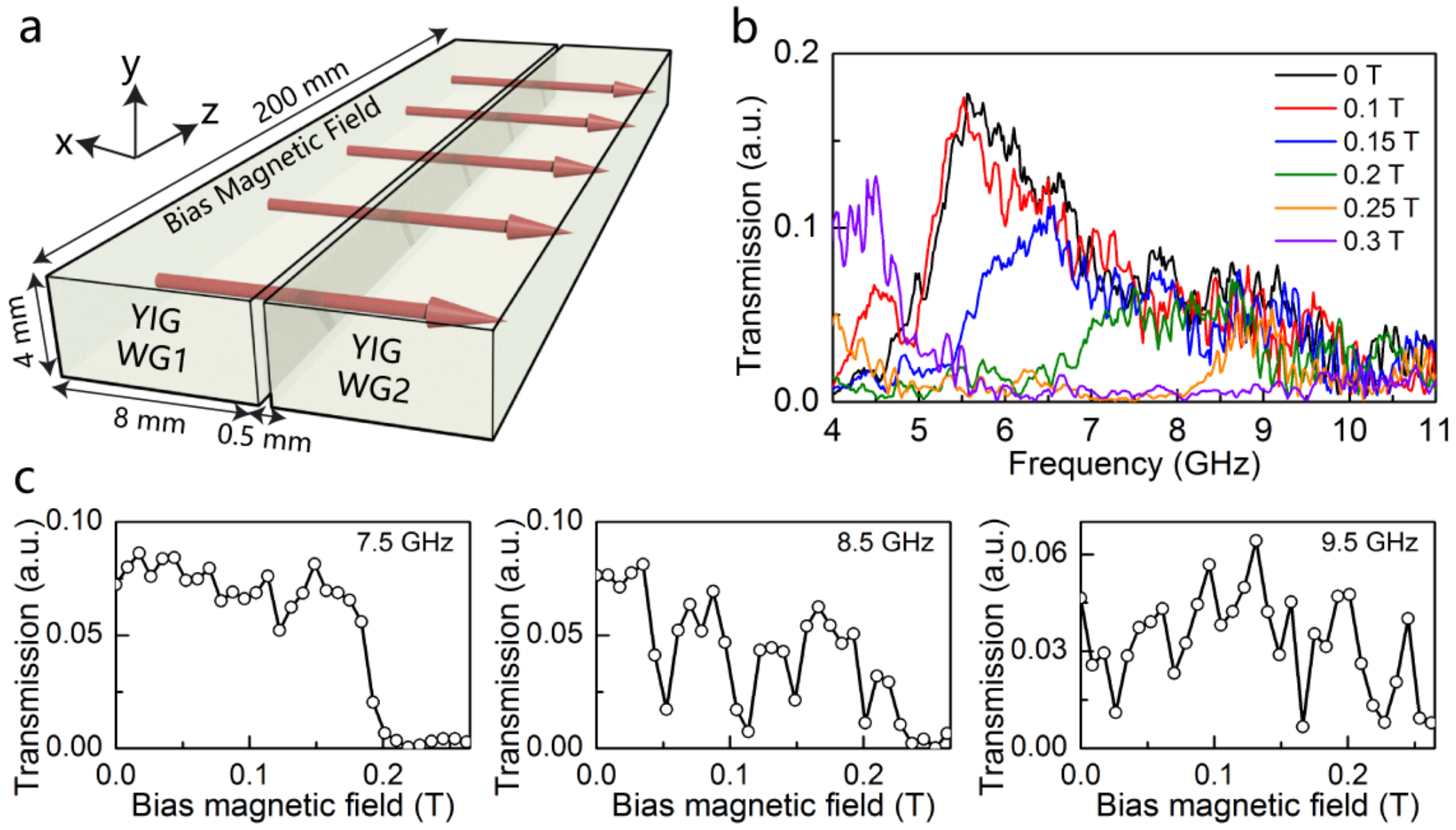

**Supplementary Figure 8 | Control experiment III. (a)** Schematic diagram of the coupled system without the microwave absorber. **(b-c)** Measured transmission spectra of this system with different bias fields.

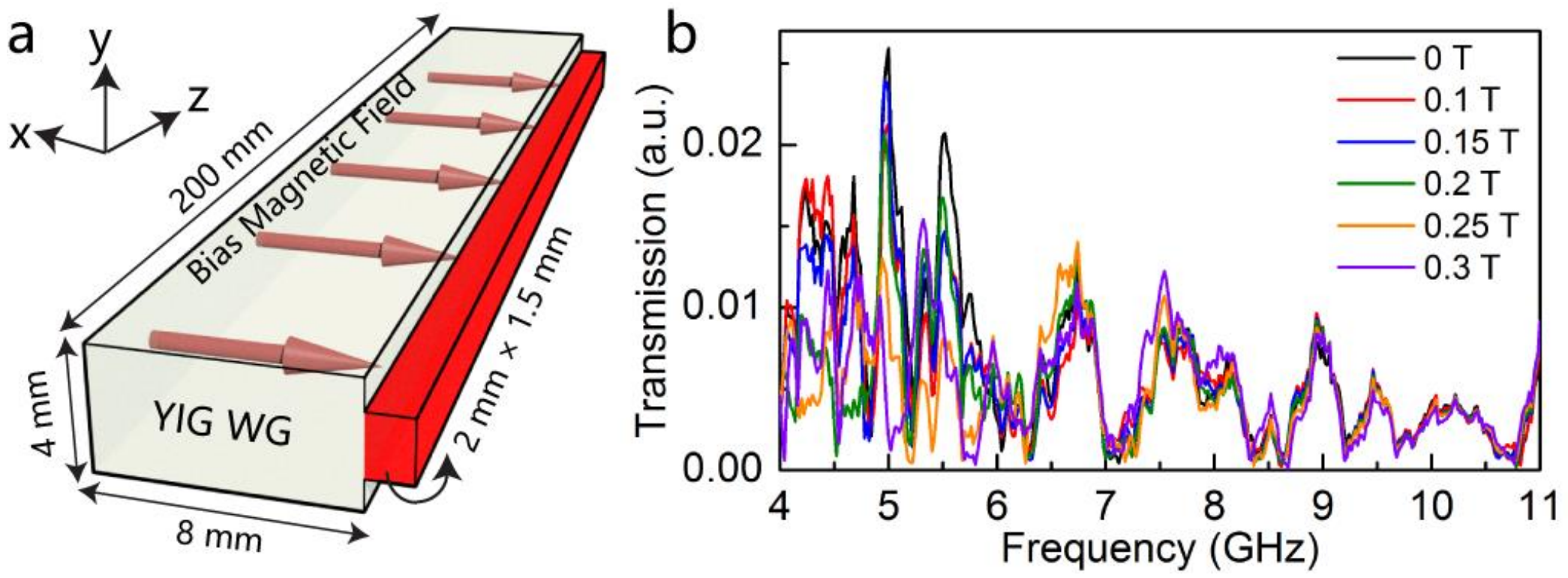

**Supplementary Figure 9 | Control experiment IV. (a)** Schematic diagram of a single YIG waveguide with the microwave absorber. **(b)** Measured transmission spectra of this system with different bias fields.

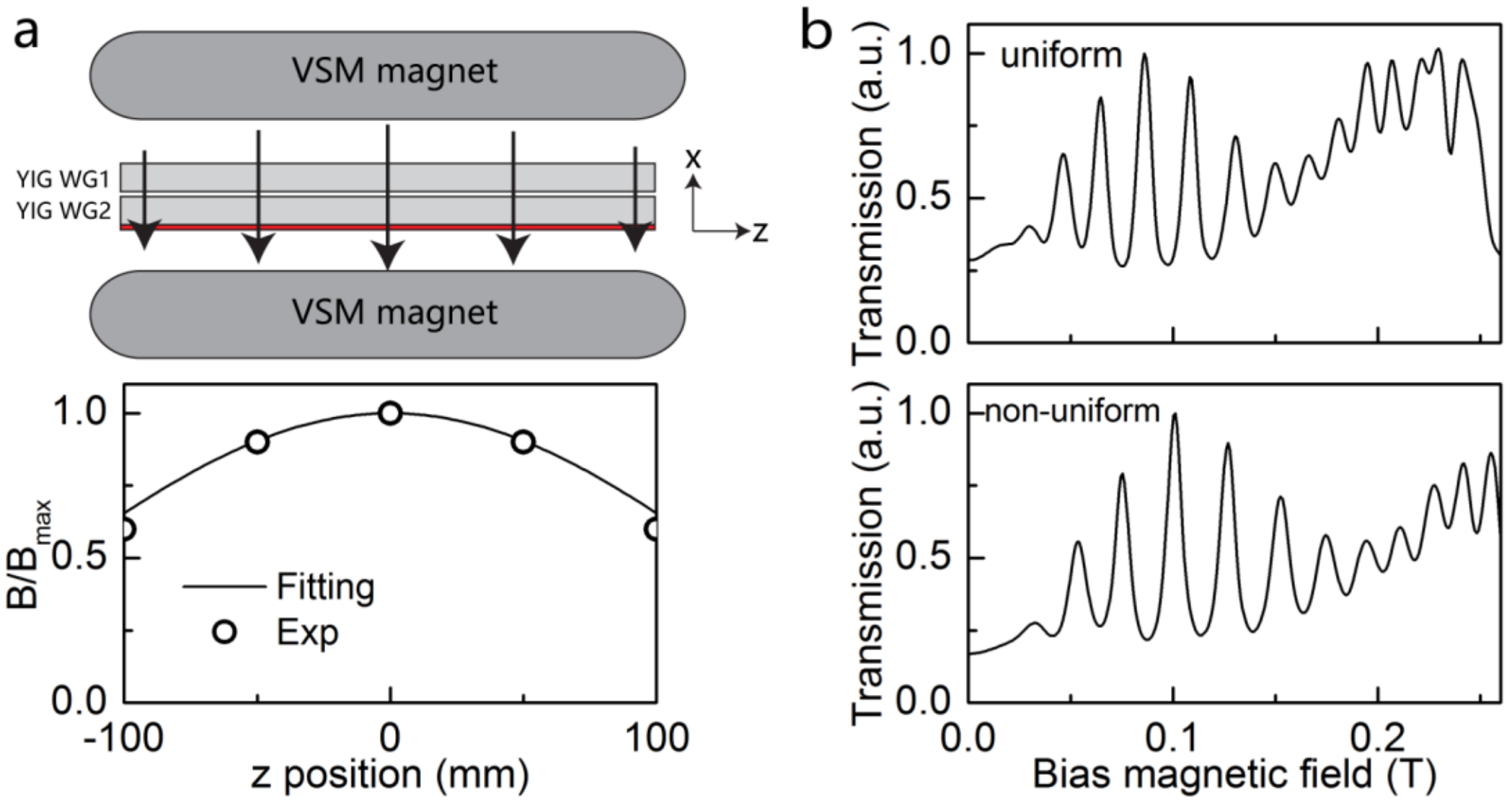

**Supplementary Figure 10 | Non-uniformity effect of the bias field. (a)** Non-uniformity of the bias magnetic field acting on the system. The circles are experimentally measure data which are fitted with the solid line for numerical simulations. **(b)** Calculated transmission spectra with uniform and non-uniform magnetic field at 9.5 GHz. We see that there is only a slight shift of the transmission spectrum if the bias magnetic field is non-uniform in a 200 mm-scale but all the physics discussed in this work remain the same. In fact, this non-uniformity has been used for encircling EPs in a 400 mm long system.

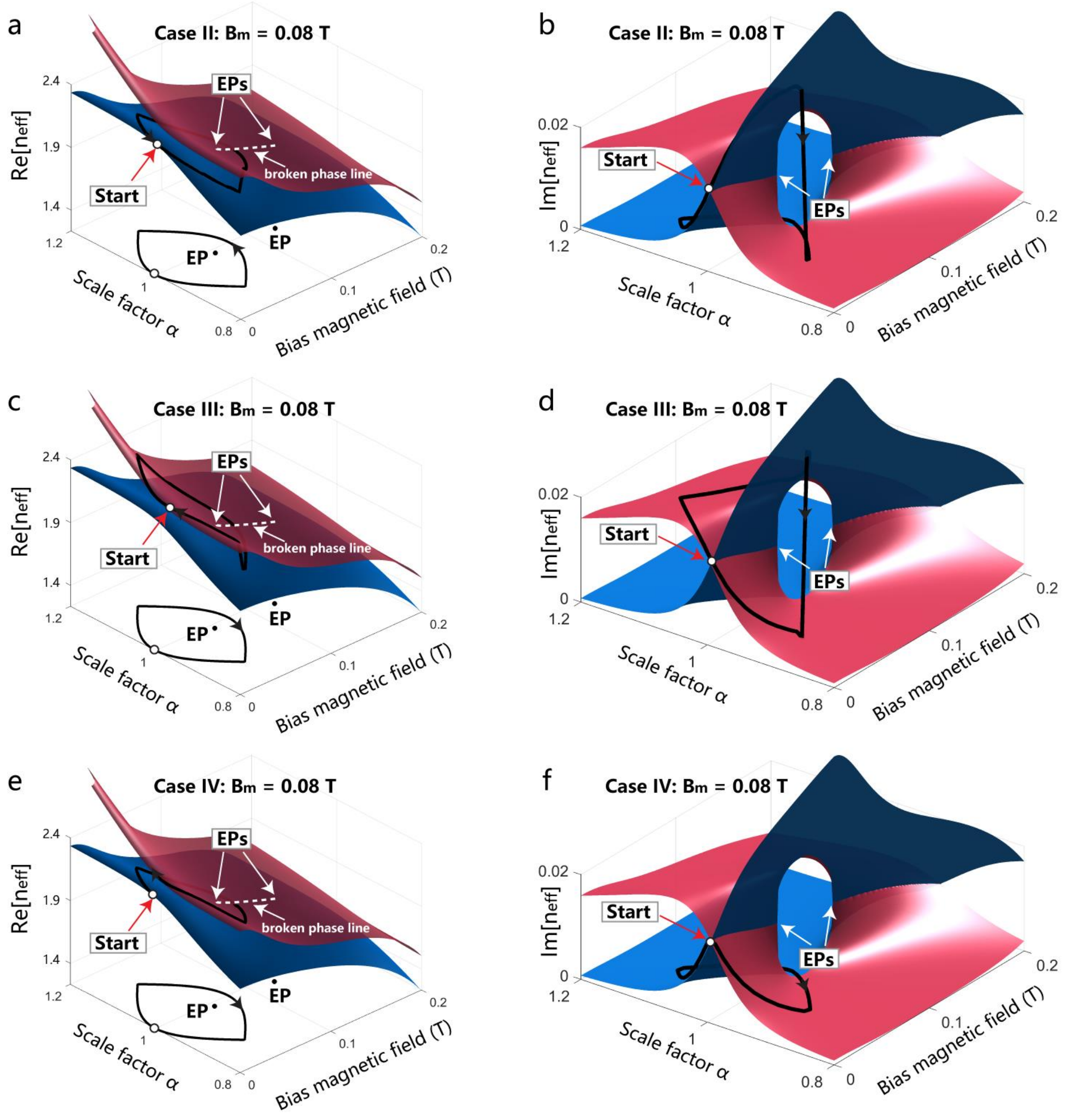

**Supplementary Figure 11 | Eigenstate evolutions on the Riemann surface for cases II-IV. (a)** Eigenvalue trajectories of a loop that encircles one EP ($B_m$ = 0.08 T) on the Riemann surface for case II **(a,b)**, case III **(c,d)** and case IV **(e,f)** (as defined in the text). The real parts are shown in **a,c,e** and the imaginary parts in **b,d,f**. A state flip occurs in case IV since the encircling takes place on the Riemann sheet with lower losses. For cases II and III, however, the encircling tends to occur on the Riemann sheet with higher losses at first but then the adiabaticity is broken down (see the sudden flip of the eigenstate in **a-d**) so that the eigenstate still returns to itself even though one EP is encircled. These trajectories are extracted from the numerical results in Fig. 6f-h of the main text.

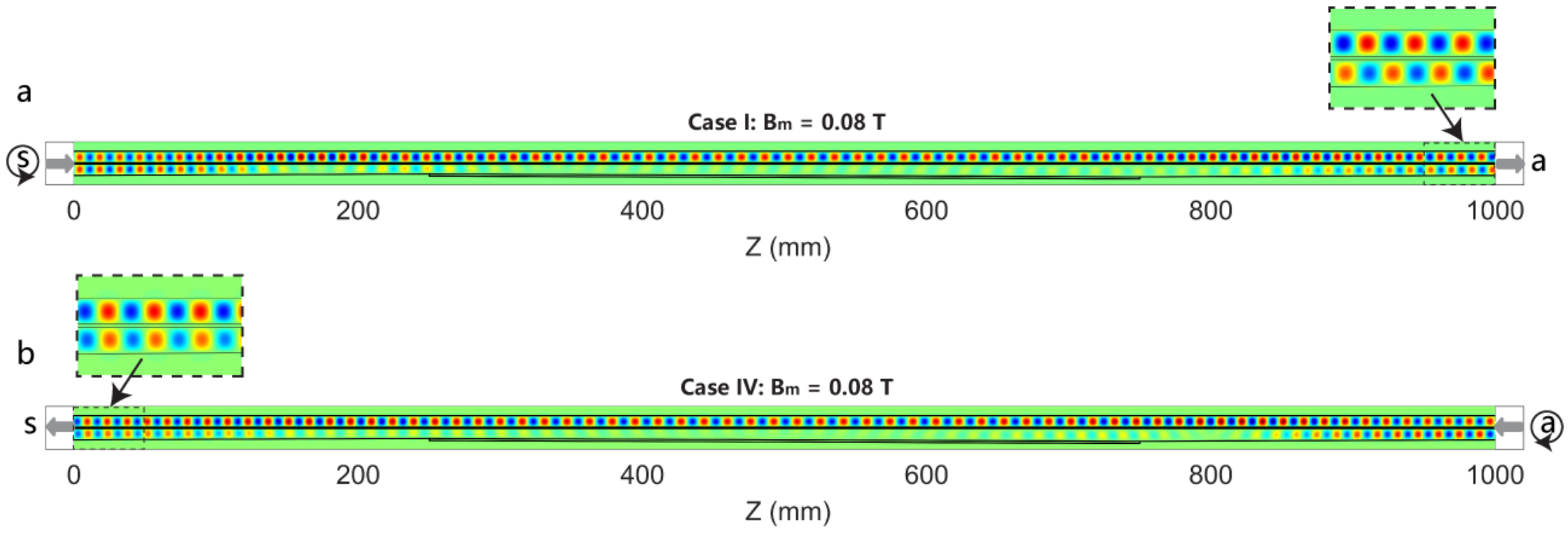

**Supplementary Figure 12 | Dynamical encircling of one EP in a 1000 mm long system.** Numerically simulated $H_y$ field distributions for case I **(a)** and case IV **(b)** with $B_m$ = 0.08 T, where the length of the system is extended to 1000 mm. In this simulation, the diameter of the VSM magnet is taken to be 500 mm, which is not the case in experiment (~200 mm). We find by comparing **a,b** with Supplementary Fig. 13d,h, the state-flip behavior is improved (i.e., the mode evolution is more adiabatic) when increasing the system length. This is however out of reach in our experiment as we do not have such a large magnet.

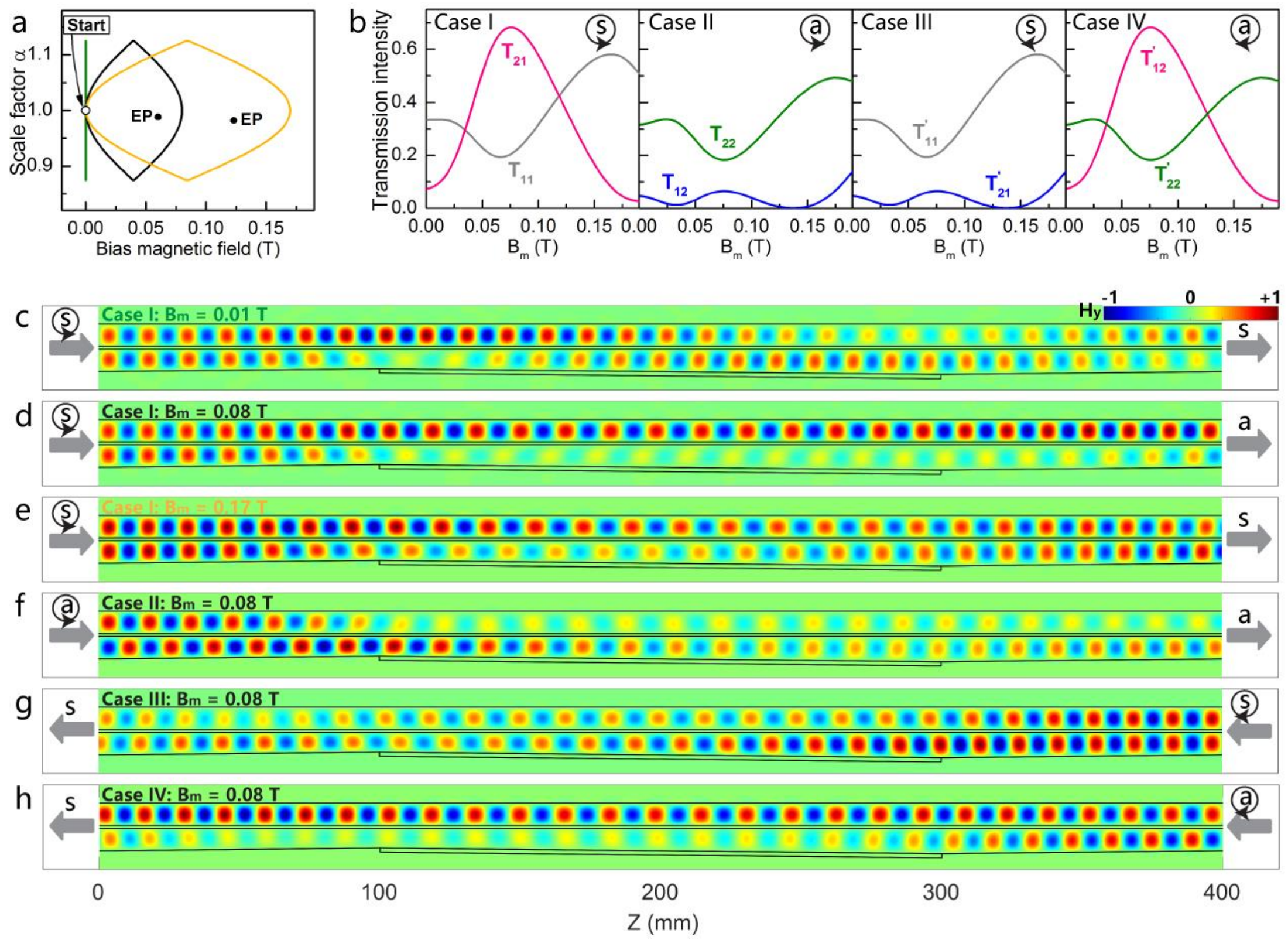

**Supplementary Figure 13 | Numerical demonstration of the dynamical encircling of EPs with half absorbers attached. (a)** Defined parameter space. **(b)** Calculated modal transmission intensities for the four cases as a function of $B_m$. **(c-h)** Numerically simulated $H_y$ field distributions for different input modes and injection directions. The items displayed here are the same as those in Fig. 6 of the main text, except that only half of YIG waveguide 2 is attached with the microwave absorber. We find that all the physics are still the same.

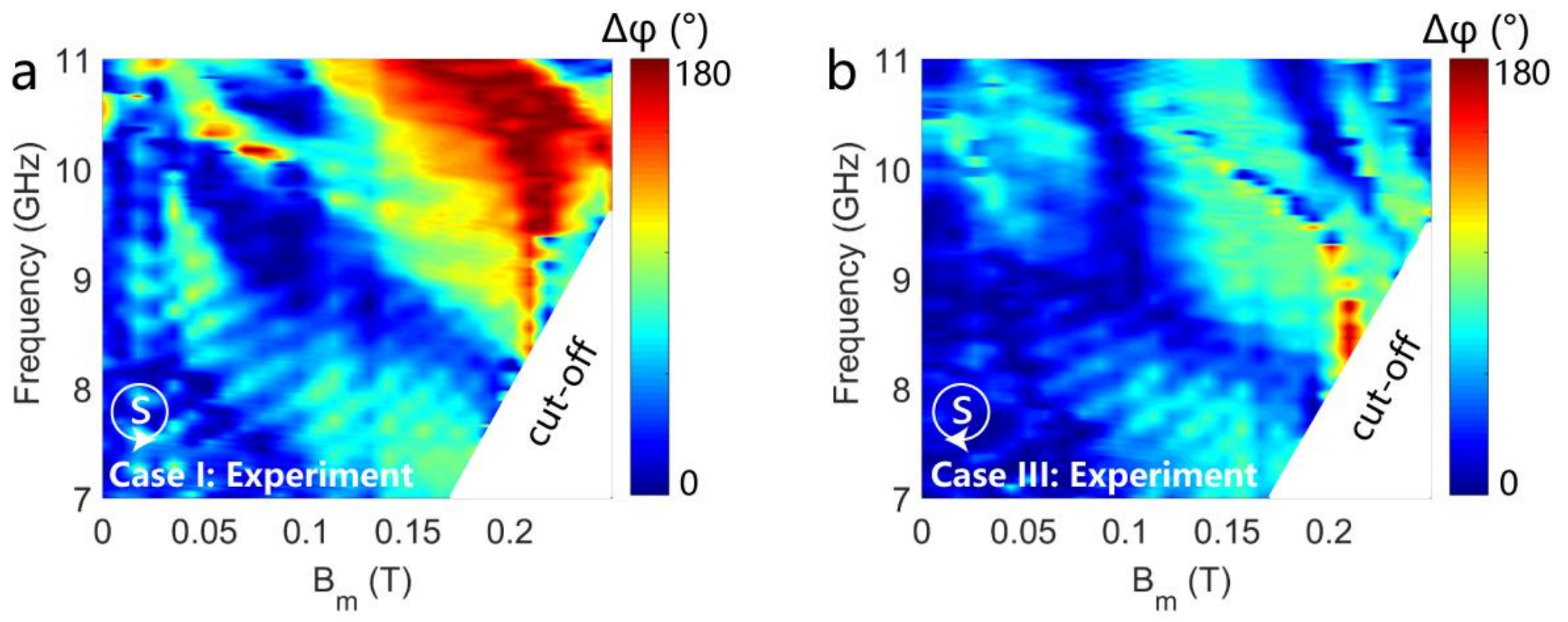

**Supplementary Figure 14 | Control experiment V.** Experimentally measured phase differences for various bias fields and frequencies for case I **(a)** and case III **(b)**, where the gap distance between the two waveguides is set to be zero.

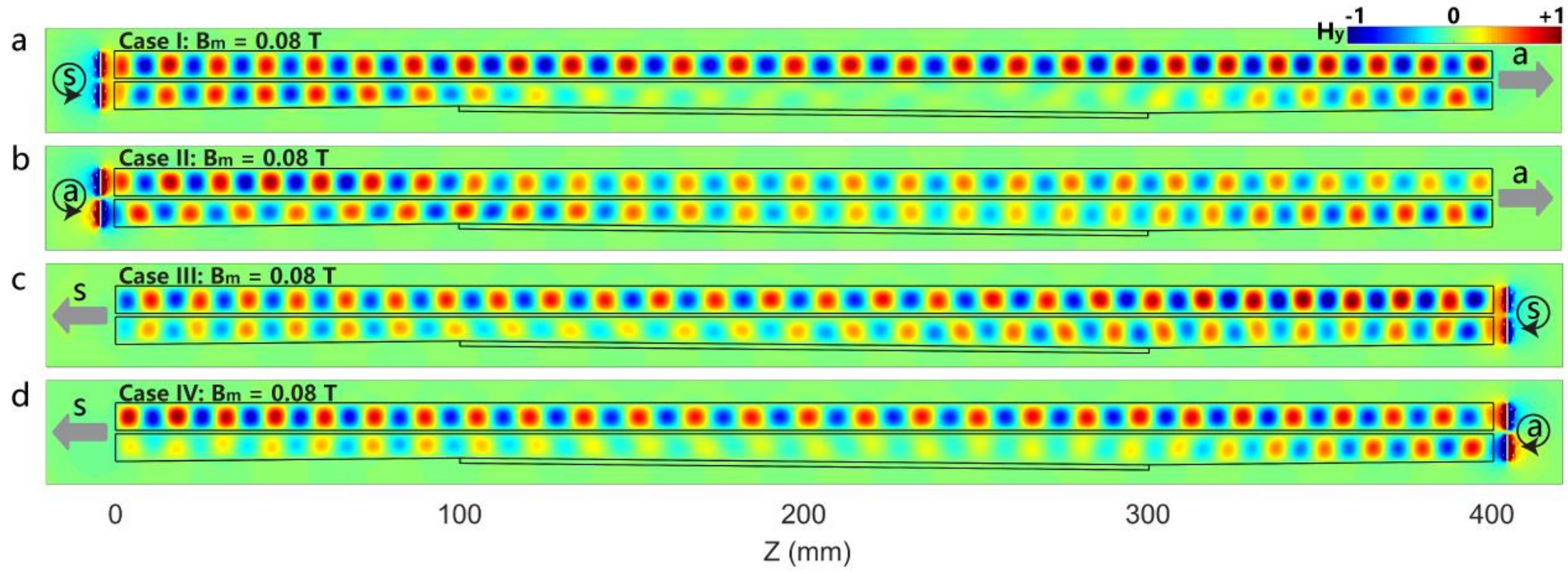

**Supplementary Figure 15 | Fabry-Perot effect when one EP is encircled. (a-d)** Numerically simulated $H_y$ field distributions for cases I-IV with $B_m$ = 0.08 T, corresponding to the encircling of one EP. The symmetric (anti-symmetric) mode is excited by two line currents with the same (opposite) oscillating directions. This is in fact the same situation as that in experiment where there is the Fabry-Perot effect. We find the field profiles in **a-d** exhibit the same features as those in Supplementary Fig. 13d,f,g,h where the Fabry-Perot effect is not considered. More specifically, a state flip occurs for cases I and IV, whereas the mode symmetry is preserved for cases II and III. Therefore, the effects arising from the encircling of EPs would not be disturbed by the Fabry-Perot effect in experiment.

# Supplementary Note 1

## Mode analysis

We give a detailed study on the origin of the degeneracy points (DPs) in the main text. We first consider the coupled ferromagnetic waveguide system shown in Fig. 1a of the main text. Supplementary Figure 1a shows the calculated effective mode index of the first four pairs of modes as a function of $\chi$. Each pair of modes consist of a symmetric and an anti-symmetric mode as a result of waveguide couplings. We find multiple DPs can be supported in the system (see the circles) due to the absence of mode couplings at specific values of $\chi$. To further explore the origin of the absence of mode couplings, we study a single ferromagnetic waveguide under the same bias magnetic field (see the inset of Supplementary Fig. 1b) and show the corresponding effective mode index as a function of $\chi$ in Supplementary Fig. 1b. The variation in the eigenfield characteristics of the single waveguide mode under a bias magnetic field should account for the origin of DPs. To show this point, we illustrate in Supplementary Fig. 1c the magnetic field distributions $H_x$ and $H_y$ in the single waveguide for the first four modes with $\chi$ = 0 (upper figures) and $\chi$ = 0.2 (lower figures). The eigenfields along the propagating $z$-axis in this paraxial waveguide ($\varepsilon_1 \approx \varepsilon_3$) are nearly zero so that they are not shown here. When the bias field is absent ($\chi$ = 0), the eigenmodes are linearly polarized, and the 1st, 2nd, 3rd and 4th mode are $E_y/H_x$, $E_x/H_y$, $E_y/H_x$ and $E_x/H_y$ polarized, respectively. The polarization of the eigenmodes becomes elliptical (as evidenced by the co-existence of $x$- and $y$-field components) when applied with a transverse bias field as shown by the results with $\chi$ = 0.2. Then we see a significant change of the eigenmode characteristics, i.e., from linear polarization to elliptical polarization upon increasing the bias magnetic field. Since the bias magnetic field in the two waveguides is along opposite direction, in the above transition process the corresponding eigenfields should change in a different way which results in the absence of mode couplings associated with DPs. The absence of mode couplings at a specific value of $\chi$ = 0.0816 can be understood from Supplementary Fig. 2, where we show the field distributions of the two uncoupled fourth set of modes, corresponding to the first DP of the fourth pair of modes in Supplementary Fig. 1a. An integration of the two uncoupled eigenfields via $\eta = \iint \left( \mathbf{E}_1 \mathbf{D}_2^* + \mathbf{H}_1 \mathbf{B}_2^* \right) d\sigma$ gives a zero coupling strength. The $z$-component of the eigenfield is very small and therefore not shown here.

We can follow the same way to demonstrate the emergence of a DP in the coupled YIG waveguide system shown in Fig. 3a of the main text. We calculate the coupling strength and show the result in Supplementary Fig. 3a. The vanishing coupling strength at ~0.095 T can account for the DP (see the inset of Fig. 3b of the main text). We further show in Supplementary Fig. 3b the eigenfield patterns of the two uncoupled modes in a single YIG waveguide at this DP. The integration between the two eigenfields via $\eta = \iint \left( \mathbf{E}_1 \mathbf{D}_2^* + \mathbf{H}_1 \mathbf{B}_2^* \right) d\sigma$ is exactly zero at this point.

## Supplementary Note 2

### Control experiments

We show the results of five control experiments in the following. Control experiments I-IV are for demonstrating the presence of an EP and a subsequent symmetry recovery, and control experiment V is for encircling the EPs in the proposed parameter space.

### Control experiment I: rotating the system by 90°

We performed a control experiment in which the coupled waveguide system in Fig. 4a of the main text is rotated by 90° around the *z*-axis, meaning that the bias field is in fact parallel to the short-edge of the YIG waveguide (see Supplementary Fig. 6a). We calculate the effective mode index as a function of the bias magnetic field at 9.5 GHz and show the results in Supplementary Fig. 6b. We find that in the field range of interest there is no EP-related physics. This is due to the fact that the system does not possess a DP when the microwave absorber is absent (see the inset of Supplementary Fig. 6b). The measured transmission spectra with different bias magnetic fields are shown in Supplementary Fig. 6c. These spectra look almost the same, indicating that no EPs or symmetry-recovery physics exist in such a configuration. This control experiment can further support our conclusion that the enhanced transmission in Fig. 4b of the main text is due to the presence of the EP and symmetry-recovery.

### Control experiment II: increasing the gap distance to 2 mm

We performed another control experiment in which the gap distance between the two YIG waveguides is increased to $g$ = 2 mm. The experimental transmission spectra are shown in Supplementary Fig. 7c, where we find the broken phase region becomes broader than that of the case $g$ = 0.5 mm (see Fig. 4b of the main text). To explain this phenomenon, we calculate the effective mode index of the lossless system at 9.5 GHz and show the results in Supplementary Fig. 7a. We note that the position of the DP exhibits a shift compared to the case of $g$ = 0.5 mm. In other words, the DP will appear at a lower bias magnetic field when the gap distance is increased. This trend is also shown in Supplementary Fig. 7d for different frequencies (comparing the red line and black line). The shift of the DP results in a shift of the broken phase region in the lossy system. We plot in Supplementary Fig. 7b the calculated imaginary part of the effective mode index of the lossy system. It is then not surprising to find that with a larger gap distance of 2 mm, the system with a zero-bias field at 9.5 GHz is already in the broken phase region (also see Supplementary Fig. 7c). The mode coupling becomes weaker when the gap distance is increased. As a consequence, with the same microwave absorber attached, the bandwidth of the broken phase region becomes broader (see Supplementary Fig. 7b for theoretical results, Supplementary Fig. 7c for experimental phenomena and Fig. 2a of the main text for interpretation).

**Control experiment III: removing the microwave absorber**

We performed a third control experiment in which the microwave absorber is removed, as illustrated schematically in Supplementary Fig. 8a. The aim of this control experiment is to demonstrate that the observed transmission enhancement phenomenon is not simply due to a coupling effect between the two lossless waveguides. Supplementary Figure 8b shows the measured transmission spectra of this system under different bias magnetic fields. Supplementary Figure 8c shows the results for three typical frequencies. As expected, we do not observe enhanced transmissions, indicating the important role of the microwave absorber since EPs can only appear in non-Hermitian systems.

**Control experiment IV: single YIG waveguide with the microwave absorber**

We performed a fourth control experiment by studying a single YIG waveguide with the microwave absorber, as illustrated schematically in Supplementary Fig. 9a. The aim of this control experiment is to demonstrate that the observed transmission enhancement phenomenon is not simply due to the variation of mode losses when the bias magnetic field varies. Supplementary Figure 9b shows the measured transmission spectra of this system with different bias magnetic fields. We found the transmission is almost unchanged because there is no other route, e.g., another lossless waveguide in the broken phase region, for the microwave to propagate through the system.

**Control experiment V: encircling the EPs without the gap**

We performed a control experiment for encircling the EPs. The gap distance is chosen to be nearly zero so that the EPs appear at a larger bias field for each frequency as discussed above. The measured phase differences for cases I and III (injected with a symmetric mode) are shown in Supplementary Fig. 14a,b, respectively. By comparing Supplementary Fig. 14a with Fig. 7b of the main text where the gap distance is ~0.5 mm, we find the state-flip region indeed moves towards larger bias fields. The result for case III here (Supplementary Fig. 14b) shows a similar behavior to that in Fig. 7c of the main text since the eigenmode exchange does not occur in case III.